\documentclass{edbk} 
\usepackage{edbkps}

\usepackage[sectionbib]{natbib}

\usepackage{epsf}

\let\footnote\savefootnote

\let\footnotetext\savefootnotetext 

\let\footnoterule\savefootnoterule 

\newcommand{\ie}{{\em i.e.}\ }
\newcommand{\eg}{{\em e.g.},\ }

\newcommand{\cf}{{\em cf.}\ }
\newcommand{\etc}{{\em etc.}\ }
\newcommand{\viz}{{\em viz.}}

\newcommand{\arcsec}{{\em ''}}

\newcommand{\twiddle}{\footnotesize $\sim$}

\def\eps@scaling{.95}
\def\epsscale#1{\gdef\eps@scaling{#1}}
\def\plotone#1{\centering \leavevmode
\epsfxsize=\eps@scaling\columnwidth \epsfbox{#1}}
\def\plottwo#1#2{\centering \leavevmode
\epsfxsize=.45\columnwidth \epsfbox{#1} \hfil
\epsfxsize=.45\columnwidth \epsfbox{#2}}
\def\plotfiddle#1#2#3#4#5#6#7{\centering \leavevmode
\vbox to#2{\rule{0pt}{#2}}
\includegraphics{#1}}

\begin{document}

\articletitle[Massive Datasets in Astronomy]{MASSIVE DATASETS IN ASTRONOMY}

\author{Robert J. Brunner, \\
S. George Djorgovski, \\
and Thomas A. Prince}
\affil{California Institute of Technology\\
Pasadena, CA 91125 USA}
\email{rb@astro.caltech.edu}
\email{george@astro.caltech.edu}
\email{prince@srl.caltech.edu}

\author{Alex S. Szalay}
\affil{Johns Hopkins University\\
Baltimore, MD 21218 USA}
\email{szalay@pha.jhu.edu}

\begin{abstract}

Astronomy has a long history of acquiring, systematizing, and
interpreting large quantities of data.  Starting from the earliest sky
atlases through the first major photographic sky surveys of the 20th
century, this tradition is continuing today, and at an ever increasing
rate.

Like many other fields, astronomy has become a very data-rich science,
driven by the advances in telescope, detector, and computer
technology.  Numerous large digital sky surveys and archives already
exist, with information content measured in multiple Terabytes, and
even larger, multi-Petabyte data sets are on the horizon.  Systematic
observations of the sky, over a range of wavelengths, are becoming the
primary source of astronomical data.  Numerical simulations are also
producing comparable volumes of information. Data mining promises to
both make the scientific utilization of these data sets more effective
and more complete, and to open completely new avenues of astronomical
research.

Technological problems range from the issues of database design and
federation, to data mining and advanced visualization, leading to a
new toolkit for astronomical research.  This is similar to challenges
encountered in other data-intensive fields today.

These advances are now being organized through a concept of the
Virtual Observatories, federations of data archives and services
representing a new information infrastructure for astronomy of the
$21^{st}$ century.  In this article, we provide an overview of some of the
major datasets in astronomy, discuss different techniques used for
archiving data, and conclude with a discussion of the future of
massive datasets in astronomy.

\end{abstract}


\section{Introduction: The New Data-Rich Astronomy}

A major paradigm shift is now taking place in astronomy and space
science.  Astronomy has suddenly become an immensely data-rich field,
with numerous digital sky surveys across a range of wavelengths, with
many Terabytes of pixels and with billions of detected sources, often
with tens of measured parameters for each object.  This is a great
change from the past, when often a single object or a small sample of
objects were used in individual studies.  Instead, we can now map the
universe systematically, and in a panchromatic manner.  This will
enable \emph{quantitatively and qualitatively new science}, from
statistical studies of our Galaxy and the large-scale structure in the
universe, to the discoveries of rare, unusual, or even completely new
types of astronomical objects and phenomena. This new digital sky,
data-mining astronomy will also enable and empower scientists and
students anywhere, without an access to large telescopes, to do
first-rate science.  This can only invigorate the field, as it opens
the access to unprecedented amounts of data to a fresh pool of talent.

Handling and exploring these vast new data volumes, and actually
making real scientific discoveries poses a considerable technical
challenge.  The traditional astronomical data analysis methods are
inadequate to cope with this sudden increase in the data volume (by
several orders of magnitude). These problems are common to all
data-intensive fields today, and indeed we expect that some of the
products and experiences from this work would find uses in other areas
of science and technology.  As a testbed for these software
technologies, astronomy provides a number of benefits: the size and
complexity of the data sets are nontrivial but manageable, the data
generally are in the public-domain, and the knowledge gained by
understanding this data is of broad public appeal.

In this chapter, we provide an overview of the state of massive
datasets in astronomy as of mid-2001. In Section~\ref{astrodata}, we briefly discuss
the nature of astronomical data, with an emphasis on understanding the
inherent complexity of data in the field. In Section~\ref{dataset}, we
present overviews of many of the largest datasets, including a
discussion of how the data are utilized and
archived. Section~\ref{vo-initaive} provides a thorough discussion of
the virtual observatory initiative, which aims to federate all of the
distributed datasets described in Section~\ref{dataset} into a
coherent archival framework. We conclude this chapter with a summary
of the current state of massive datasets in astronomy.

\section{The Nature of Astronomical Data\label{astrodata}}

By its inherent nature, astronomical data are extremely heterogeneous,
in both format and content. Astronomers are now exploring all regions
of the electromagnetic spectrum, from gamma-rays through radio
wavelengths. With the advent of new facilities, previously unexplored
domains in the gravitational spectrum will soon be available, and
exciting work in the astro-particle domain is beginning to shed light
on our Universe. Computational advances have enabled detailed physical
simulations which rival the largest observational datasets in terms of
complexity. In order to truly understand our cosmos, we need to
assimilate all of this data, each presenting its own physical view of
the Universe, and requiring its own technology.

\begin{figure}[!hbt]
\plotfiddle{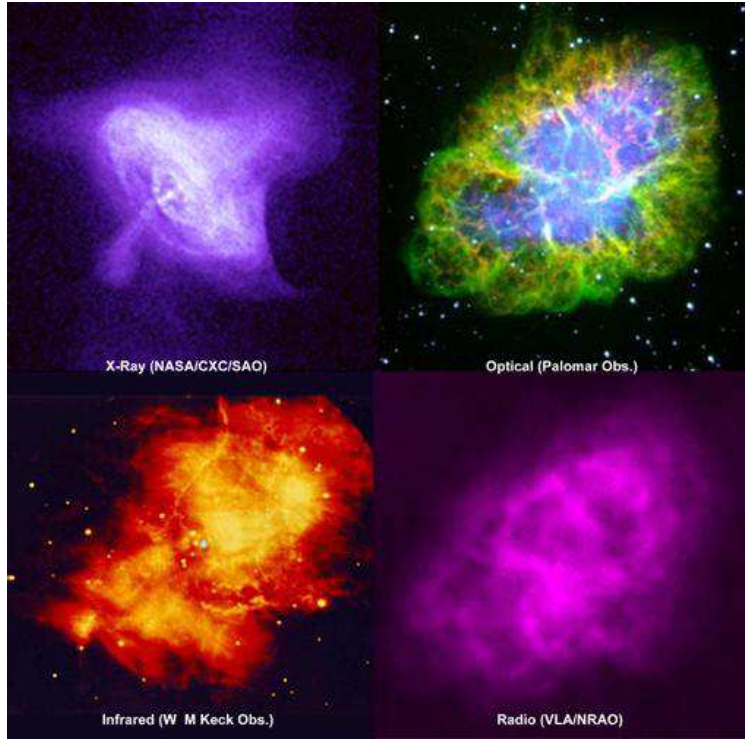}{4.0in}{0}{50}{50}{-150}{0}
\caption{A multiwavelength view of the Crab nebula, a supernova 
remnant that was first sighted by Chinese astronomers in 1054
AD. Clearly demonstrated in this montage of images are the different
physical processes that are manifested in the different spectral
regions. Image Credit: NASA/CXC/SAO.\label{crab}}
\end{figure}

Despite all of this heterogeneity, however, astronomical data and its
subsequent analysis can be broadly classified into five domains. In
order to clarify later discussions, we briefly discuss these domains
and define some key astrophysical concepts which will be utilized
frequently throughout this chapter.

\begin{itemize}

\item Imaging data is the fundamental constituent of astronomical observations, 
capturing a two-dimensional spatial picture of the Universe within a
narrow wavelength region at a particular epoch or instant of
time. While this may seem obvious to most people---after all, who
hasn't seen a photograph---astrophysical pictures (see, \eg
Figures~\ref{crab} and~\ref{image}) are generally taken through a specific filter, or
with an instrument covering a limited range of the electromagnetic
spectrum, which defines the wavelength region of the
observation. Astronomical images can be acquired directly, \eg with
imaging arrays such as CCDs\footnote{Charge Coupled Device---a digital
photon counting device that is superior to photographic images in both
the linearity of their response and quantum efficiency}, or
synthesized from interferometric observations as is customarily done
in radio astronomy.

\begin{figure}[!hbt]

\plottwo{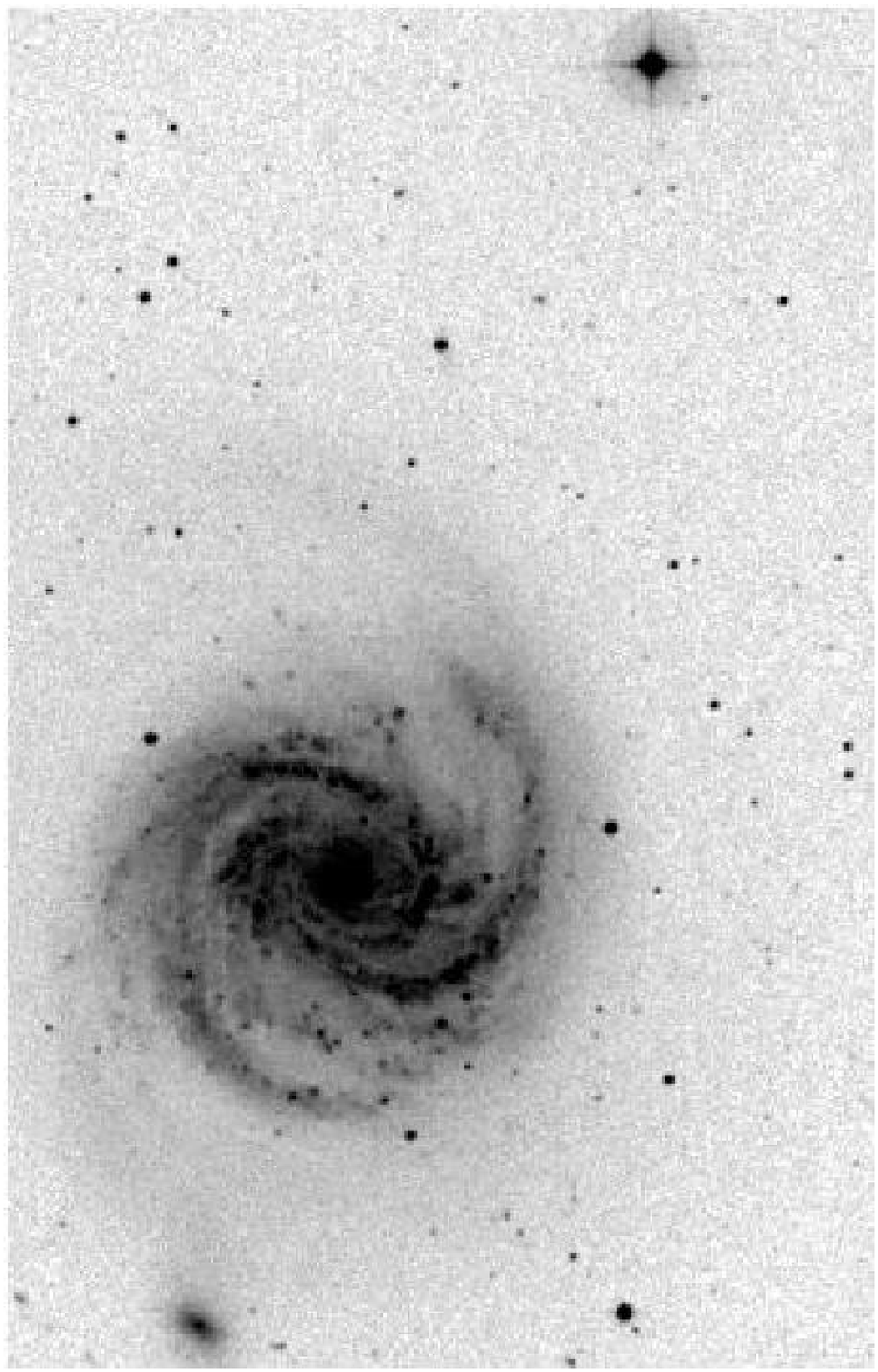}{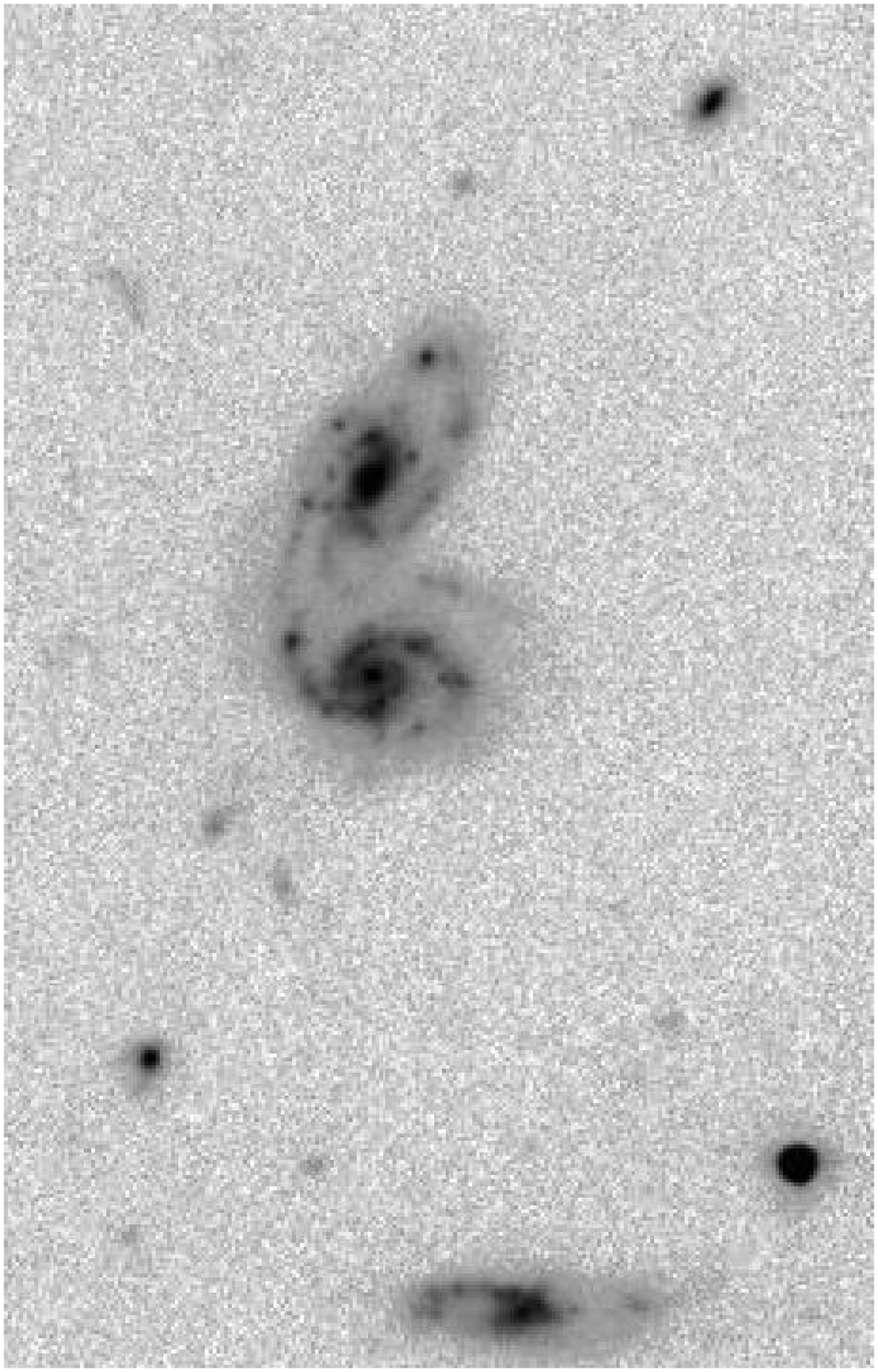}
\caption{Making the tradeoff between area and resolution. The image on 
the left is from the ground-based DPOSS survey (see below) of the
field of M100, a nearby spiral galaxy. While the entire survey covers
on-half of the entire sky, this single image is only one-millionth of
the size of the entire sky (\ie one microsky). The image on the right
is a subset from the deepest optical image ever taken, the STIS clear
image of the Hubble Deep Field South, image credit: R. Williams
(STScI), the HDF-S Team, and NASA. This image is $10,000$ times
smaller than the DPOSS image, thus representing 100 picosky.
\label{image}}
\end{figure}

\item Catalogs are generated by processing the imaging data. Each detected source can 
have a large number of measured parameters, including coordinates,
various flux quantities, morphological information, and areal
extant. In order to be detected, a source must stand out from the
background noise (which can be either cosmic or instrumental in
origin).  The significance of a detection is generally quoted in terms
of $\sigma$, which is a relative measure of the strength of the source
signal relative to the dispersion in the background noise. We note
that the source detection process is generally limited both in terms
of the flux (total signal over the background) and surface brightness
(intensity contrast relative to the background).

Coordinates are used to specify the location of astronomical sources
in the sky. While this might seem obvious, the fact that we are sited
in a non-stationary reference frame (\eg the earth rotates, revolves
around the sun, and the sun revolves around the center of our Galaxy)
complicates the quantification of a coordinate location. In addition,
the Earth's polar axis precesses, introducing a further
complication. As a result, coordinate systems, like Equatorial
coordinates, must be fixed at a particular instant of time (or epoch),
to which the actual observations, which are made at different times,
can be transformed. One of the most popular coordinate systems is
J2000 Equatorial, which is fixed to the initial instant (zero hours
universal time) of January 1, 2000. One final caveat is that nearby
objects (\eg solar system bodies or nearby stars) move on measurable
timescales. Thus the date or precise time of a given observations must
also be recorded.

Flux quantities determine the amount of energy that is being received
from a particular source. Since different physical processes emit
radiation at different wavelengths, most astronomical images are
obtained through specific filters. The specific filter(s) used varies,
depending on the primary purpose of the observations and the type of
recording device. Historically, photographic surveys used filters
which were well matched to the photographic material, and have names
like $O$, $E$, $J$, $F$, and $N$. More modern digital detectors have
different characteristics (including much higher sensitivity), and
work primarily with different filter systems, which have names like
$U$, $B$, $V$, $R$, and $I$, or $g$, $r$, $i$ in the optical, and $J$,
$H$, $K$, $L$, $M$, and $N$ in the near-infrared.

In the optical and infrared regimes, the flux is measured in units of
magnitudes (which is essentially a logarithmic rescaling of the
measured flux) with one magnitude equivalent to $-4$ decibels. This is
the result of the historical fact that the human eye is essentially a
logarithmic detector, and astronomical observations have been made and
recorded for many centuries by our ancestors. The zeropoint of the
magnitude scale is determined by the star Vega, and thus all flux
measurements are relative to the absolute flux measurements of this
star. Measured flux values in a particular filter are indicated as $B
= 23$ magnitudes, which means the measured $B$ band flux is $10^{0.4
\times 23}$ times fainter than the star Vega in this band. At other
wavelengths, like x-ray and radio, the flux is generally quantified in
standard physical units such as ergs cm$^{-2}$ s$^{-1}$ Hz$^{-1}$. In
the radio, observations often include not only the total intensity
(indicated by the Stokes $I$ parameter), but also the linear
polarization parameters (indicated by the Stokes $Q$, and $U$
parameters).

\item Spectroscopy, Polarization, and other follow-up measurements provide detailed physical 
quantification of the target systems, including distance information
(\eg redshift, denoted by $z$ for extragalactic objects), chemical
composition (quantified in terms of abundances of heavier elements
relative to hydrogen), and measurements of the physical (\eg
electromagnetic, or gravitational) fields present at the source. An
example spectrum is presented in Figure~\ref{spectra}, which also
shows the three optical filters used in the DPOSS survey (see below)
superimposed.

\begin{figure}[!hbt]
\plotfiddle{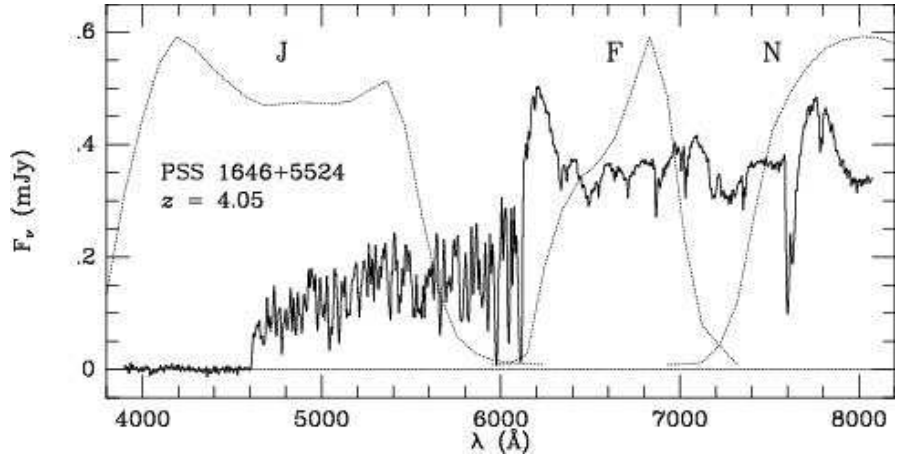}{2.25in}{0}{60}{60}{-175}{-18}
\caption{A spectrum of a typical $z > 4$ quasar PSS 1646+5524, 
with the DPOSS photographic filter transmission curves ($J$, $F$, and
$N$) overplotted as dotted lines.  The prominent break in the spectrum
around an observed wavelength of $6000$ Angstroms is caused by absorption by
intergalactic material (that is material between us and the quasar)
that is referred to as the Ly$\alpha$ forest. The redshift of this
source can be calculated by knowing that this absorption occurs for
photons more energetic than the Ly$\alpha$ line which is emitted at
$1216$ Angstroms.\label{spectra}}
\end{figure}

\item Studying the time domain (see, \eg Figure~\ref{time}) provides 
important insights into the nature of the Universe, by identifying
moving objects (\eg near-Earth objects, and comets), variable sources
(\eg pulsating stars), or transient objects (\eg supernovae, and
gamma-ray bursts). Studies in the time domain either require multiple
epoch observations of fields (which is possible in the overlap regions
of surveys), or a dedicated synoptic survey. In either case, the data
volume, and thus the difficulty in handling and analyzing the
resulting data, increases significantly.

\begin{figure}[!hbt]
\plotfiddle{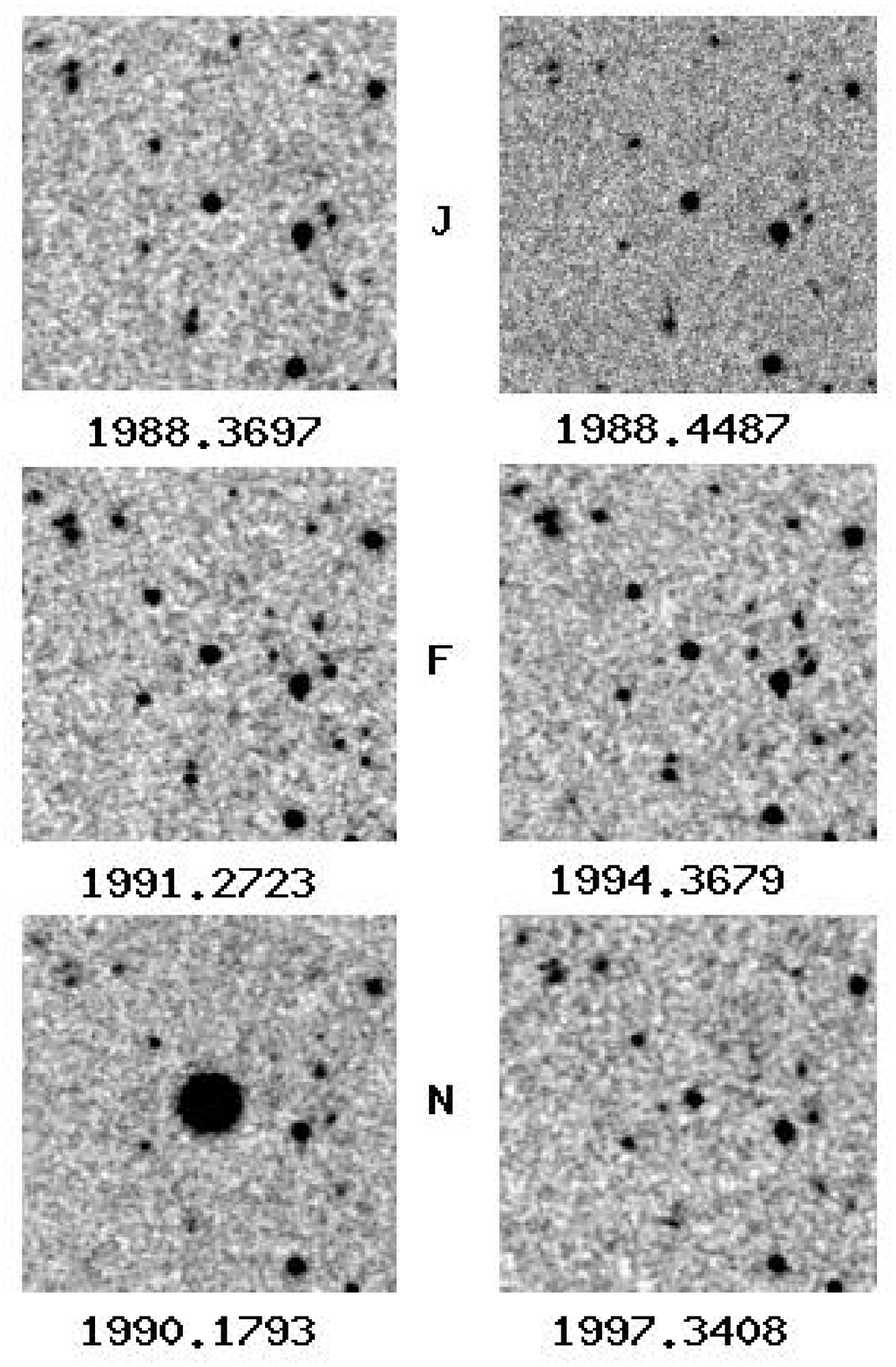}{3.85in}{0}{40}{40}{-125}{-18}
\caption{Example of a discovery in the time domain. Images of a star, 
PVO 1558+3725, seen in the DPOSS plate overlaps in $J$ (= green, top)
$F$ (= red, middle) and $N$ ($\approx $ near-infrared, bottom).  Since
the plates for the POSS-II survey were taken at different epochs (\ie
they were taken on different days), that can be separated by several
years (the actual observational epoch is indicated below each panel),
we have a temporal recording of the intensity of the star.  Notice how
the central star is significantly brighter in the lower right
panel. Subsequent analysis has not indicated any unusual features,
and, as a result, the cause, amplitude, and duration of the outburst
are unknown.\label{time}}
\end{figure}

\item Numerical Simulations are theoretical tools which can be compared with 
observational data. Examples include simulations of the formation and
evolution of large-scale structure in the Universe, star formation in
our Galaxy, supernova explosions, \etc Since we only have one Universe
and cannot modify the initial conditions, simulations provide a
valuable tool in understanding how the Universe and its constituents
formed and have evolved. In addition, many of the physical processes
that are involved in these studies are inherently complex. Thus direct
analytic solutions are often not feasible, and numerical analysis is
required. 

\end{itemize}

\section{Large Astronomical Datasets\label{dataset}}

As demonstrated below, there is currently a great deal of archived
data in Astronomy at a variety of locations in a variety of different
database systems systems. In this section we focus on ground-based
surveys, ground-based observatories, and space-based observatories. We
do not include any discussion of information repositories such as the
Astrophysics Data System\footnote{http://adswww.harvard.edu/} (ADS),
the Set of Identifications, Measurements, and Bibliography for
Astronomical Data\footnote{http://cdsweb.u-strasbg.fr/Simbad.html}
(SIMBAD), or the NASA Extragalactic
Database\footnote{http://ned.ipac.caltech.edu} (NED), extremely
valuable as they are. This review focuses on more homogeneous
collections of data from digital sky surveys and specific missions
rather than archives which are more appropriately described as digital
libraries for astronomy.

Furthermore, we do not discuss the large number of new initiatives,
including the Large-Aperture Synoptic Survey Telescope (LSST), the
California Extremely Large Telescope (CELT), the Visible and Infrared
Survey Telescope\footnote{http://www.vista.ac.uk} (VISTA), or the Next
Generation Space Telescope\footnote{http://www.ngst.stsci.edu/}
(NGST), which will provide vast increases in the quality and quantity
of astronomical data.

\subsection{Ground-Based Sky Surveys}

Of all of the different astronomical sources of data, digital sky
surveys are the major drivers behind the fundamental changes underway
in the field. Primarily this is the result of two factors: first, the
sheer quantity of data being generated over multiple wavelengths, and
second, as a result of the homogeneity of the data within each
survey. The federation of different surveys would further improve the
efficacy of future ground- and space-based targeted observations, and
also open up entirely new avenues for research.

In this chapter, we describe only some of the currently existing
astronomical archives as examples of the types, richness, and quantity
of astronomical data which is already available. Due to the space
limitations, we cannot cover many other, valuable and useful surveys,
experiments and archives, and we apologize for any omissions. This
summary is not meant to be complete, but merely illusory.

Photographic plates have long endured as efficient mechanisms for
recording surveys (they have useful lifetimes in excess of one hundred
years and offer superb information storage capacity, but unfortunately
they are not directly computer-accessible and must be digitized before
being put to a modern scientific use). Their preeminence in a digital
world, however, is being challenged by new technologies. While many
photographic surveys have been performed, \eg from the Palomar Schmidt
telescope\footnote{A Schmidt telescope has an optical design which
allows it to image a very wide field, typically several degrees on a
side. This is in contrast to most large telescopes which have field of
views that are measured in arcminutes.} in California, and the UK
Schmidt telescope in New South Wales, Australia, these data become
most useful when the plates are digitized and cataloged.

While we describe two specific projects, as examples, several other
groups have digitized photographic surveys and generated and archived
the resulting catalogs, including the Minnesota Automated Plate
Scanner\footnote{http://aps.umn.edu/}~\citep[APS;][]{pennington93},
the Automated Plate Measuring
Machine\footnote{http://www.ast.cam.ac.uk/\twiddle
mike/casu/apm/apm.html}~\citep[APM;][]{mcmahon92} at the Institute of
Astronomy, Cambridge, UK, the coordinates, sizes, magnitudes,
orientations, and shapes~\citep[COSMOS;][]{yentis92} and its
successor,
SuperCOSMOS\footnote{http://www.roe.ac.uk/cosmos/scosmos.html}, plate
scanning machines at the Royal Observatory Edinburgh. Probably the
most popular of the digitized sky surveys (DSS) are those produced at
the Space Telescope Science
Institute\footnote{http://archive.stsci.edu/dss/} (STScI) and its
mirror sites in Canada\footnote{http://cadcwww.dao.nrc.ca/dss/},
Europe\footnote{http://archive.eso.org/dss/dss/}, and
Japan\footnote{http://dss.nao.ac.jp}.

\begin{description}

\item[DPOSS] The Digitized Palomar Observatory Sky Survey\footnote{http://dposs.caltech.edu/} (DPOSS)
is a digital survey of the entire Northern Sky in three visible-light
bands, formally indicated by $g$, $r$, and $i$ (blue-green, red, and
near-infrared, respectively).  It is based on the photographic sky
atlas, POSS-II, the second Palomar Observatory Sky Survey, which was
completed at the Palomar 48-inch Oschin Schmidt
Telescope~\citep{reid91}.  A set of three photographic plates, one in
each filter, each covering 36 square degrees, were taken at each of
894 pointings spaced by 5 degrees, covering the Northern sky (many of
these were repeated exposures, due to various artifacts such as the
aircraft trails, plate defects, {\em etc.}).  The plates were then digitized at the Space Telescope Science
Institute (STScI), using a laser micro-densitometer.  The plates are
scanned with 1.0\arcsec pixels, in rasters of 23,040 square, with 16
bits per pixel, producing about 1 Gigabyte per plate, or about 3
Terabytes of pixel data in total.

These scans were processed independently at STScI (for the
purposes of constructing a new guide star catalog for the HST) and at
Caltech (for the DPOSS project).  Catalogs of all the detected objects
on each plate were generated, down to the flux limit of the plates,
which roughly corresponds to the equivalent blue limiting magnitude of
approximately $22$.  A specially developed software package, called
SKICAT (Sky Image Cataloging and Analysis Tool;~\citealt{weir95}) was used
to analyze the images.  SKICAT incorporates some machine learning
techniques for object classification and measures about 40 parameters
for each object in each band.  Star-galaxy classification was done
using several methods, including decision trees and neural nets; for
brighter galaxies, a more detailed morphological classification may be
added in the near future.  The DPOSS project also includes an
extensive program of CCD calibrations done at the Palomar 60-inch
telescope. These CCD data were used both for magnitude calibrations,
and as training data sets for object classifiers in SKICAT.  The
resulting object catalogs were combined and stored in a Sybase
relational DBMS system; however, a more powerful system is currently
being implemented for more efficient scientific exploration. This
new archive will also include the actual pixel data in the form of
astrometrically and photometrically calibrated images.

The final result of DPOSS will be the Palomar Norris Sky Catalog
(PNSC), which is estimated to contain about 50 to 100 million
galaxies, and between 1 and 2 billion stars, with over 100 attributes
measured for each object, down to the equivalent blue limiting
magnitude of $22$, and with star-galaxy classifications accurate to
90\% or better down to the equivalent blue magnitude of approximately
$21$. This represents a considerable advance over other, currently
existing optical sky surveys based on large-format photographic
plates.  Once the technical and scientific verification of the final
catalog is complete, the DPOSS data will be released to the
astronomical community.

As an indication of the technological evolution in this field, the
Palomar Oschin Schmidt telescope is now being retrofitted with a large
CCD camera (QUEST2) as a collaborative project between Yale
University, Indiana University, Caltech, and JPL. This will lead to
pure digital sky surveys from Palomar Observatory in the future.

\item[USNO-A2] The United States Naval Observatory Astrometric (USNO-A2)
catalog\footnote{http://ftp.nofs.navy.mil/projects/pmm/a2.html}~\citep{monet96}
is a full-sky survey containing over five hundred million unresolved
sources down to a limiting magnitude of $B \sim 20$ whose positions
can be used for astrometric references.  These sources were detected
by the Precision Measuring Machine (PMM) built and operated by the
United States Naval Observatory Flagstaff Station during the scanning
and processing of the first Palomar Observatory Sky Survey (POSS-I) O and
E plates, the UK Science Research Council SRC-J survey plates, and the
European Southern Observatory ESO-R survey plates. The total amount of
data utilized by the survey exceeds 10 Terabytes.

The USNO-A2 catalog is provided as a series of binary files, organized
according to the position on the sky. Since the density of sources on
the sky varies (primarily due to the fact that our galaxy is a disk
dominated system), the number of sources in each file varies
tremendously. In order to actually extract source parameters, special
software, which is provided along with the data is required. The
catalog includes the source position, right ascension and declination
(in the J2000 coordinate system, with the actual epoch derived as the
mean of the blue and red plate) and the blue and red magnitude for
each star. The astrometry is tied to the ACT
catalog~\citep{urban97}. Since the PMM detects and processes at and
beyond the nominal limiting magnitude of these surveys, a large number
of spurious detections are initially included in the operational
catalog. In order to improve the efficacy of the catalog, sources were
required to be spatially coincident, within a 2\arcsec\ radius
aperture, on the blue and red survey plate.

\item[SDSS] The Sloan Digital Sky Survey\footnote{http://www.sdss.org/} (SDSS) 
is a large astronomical collaboration focused on constructing the
first CCD photometric survey of the North Galactic hemisphere (10,000
square degrees---one-fourth of the entire sky). The estimated 100
million cataloged sources from this survey will then be used as the
foundation for the largest ever spectroscopic survey of galaxies,
quasars and stars.

The full survey is expected to take five years, and has recently begun
full operations. A dedicated 2.5m telescope is specially designed to
take wide field (3 degree x 3 degree) images using a 5 by 6 mosaic of
2048x2048 CCD's, in five wavelength bands, operating in scanning
mode. The total raw data will exceed 40 TB. A processed subset, of
about 1 TB in size, will consist of 1 million spectra, positions and
image parameters for over 100 million objects, plus a mini-image
centered on each object in every color. The data will be made
available to the public (see, \eg Figure~\ref{sdss} for a public SDSS
portal) at specific release milestones, and upon completion of the
survey.

During the commissioning phase of the survey data was obtained, in
part to test out the hardware and software components. Already, a
wealth of new science has emerged from this data.

The Sloan Digital Sky Survey (SDSS) is a joint project of The
University of Chicago, Fermilab, the Institute for Advanced Study, the
Japan Participation Group, The Johns Hopkins University, the
Max-Planck-Institute for Astronomy (MPIA), the Max-Planck-Institute
for Astrophysics (MPA), New Mexico State University, Princeton
University, the United States Naval Observatory, and the University of
Washington.

\begin{figure}[!hbt]
\plotfiddle{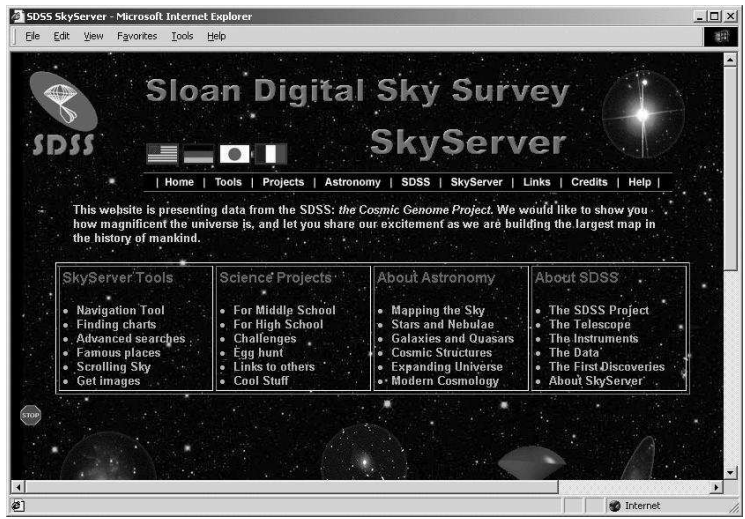}{3.0in}{0}{50}{50}{-160}{0}
\caption{A public portal into the Sloan Digital Sky Survey, 
which is the cosmic equivalent to the Human Genome project.\label{sdss}}
\end{figure}

\end{description}

There is a large number of experiments and surveys at aimed at
detecting time-variable sources or phenomena, such as gravitational
microlensing, optical flashes from cosmic gamma-ray bursts, near-Earth
asteroids and other solar system objects, \etc A good list of project
web-sites is maintained by Professor Bohdan Paczynski at
Princeton\footnote{http://astro.princeton.edu/faculty/bp.html}. Here
we describe several interesting examples of such projects.

\begin{description}

\item[MACHO] The MACHO project\footnote{http://wwwmacho.anu.edu.au/} 
was one of the pioneering astronomical projects in generating large
datasets. This project was designed to look for a particular type of
dark matter collectively classified as Massive Compact Halo Objects
(\eg brown dwarfs or planets) from whence the project's name was
derived. The signature for sources of this type is the amplification
of the light from extragalactic stars by the gravitational lens effect
of the intervening MACHO. While the amplitude of the amplification can
be large, the frequency of such events is extremely rare. Therefore,
in order to obtain a statistically useful sample, it is necessary to
photometrically monitor several million stars over a period of several
years. The MACHO Project is a collaboration between scientists at the
Mt. Stromlo and Siding Spring Observatories, the Center for Particle
Astrophysics at the Santa Barbara, San Diego, and Berkeley campuses of
the University of California, and the Lawrence Livermore National
Laboratory.

The MACHO project built a two channel system that employs eight
2048x2048 CCDs, which was mounted and operated on the 50-inch
telescope at Mt. Stromlo. This large CCD instrument presented a high
data rate (especially given that the survey commenced in 1992) of
approximately several Gigabytes per night. Over the course of the
survey, nearly 100,000 images were taken and processed, with a total
data volume exceeding 7 Terabytes. While the original research goal of
finding microlensing events was realized (essentially by a real-time
data-analysis system), the MACHO data provides an enormously useful
resource for studying a variety of variable sources. Unfortunately,
funding was never secured to build a data archive, limiting the
utility of the data primarily to only those members of the MACHO
science team. Another similar project is the
OGLE-II\footnote{http://astro.princeton.edu/\twiddle ogle}, or the
second Optical Gravitational Lensing Experiment, which has a total
data volume in excess of one Terabyte.

\item[ROTSE] The Robotic Optical Transient Search 
Experiment\footnote{http://www.umich.edu/\twiddle rotse} is an experimental
program to search for astrophysical optical transients on time scales
of a fraction of a second to a few hours. While the primary incentive
of this experiment has been to find the optical counterparts of
gamma-ray bursts (GRBs), additional variability studies have been
enabled, including a search for orphan GRB afterglows, and an analysis
of a particular type of variable star, known as an RR Lyrae, which
provides information on the structure of our Galaxy.

The ROTSE project initially began operating in 1998, with a four-fold
telephoto array, imaging the whole visible sky twice a night to
limiting flux limit of approximately $15.5$. The total data volume for
the original project is approximately four Terabytes. Unlike other
imaging programs, however, the large field of view of the telescope
results in a large number of sources per field (approximately
40,000). Therefore, reduction of the imaging data to object lists does
not compress the data as much as is usual in astronomical data. The
data is persisted on a robotic tape library, but insufficient
resources have prevented the creation of a public archive.

The next stage of the ROTSE project is a set of four (and eventually
six) half meter telescopes to be sited globally. Each telescope has a
2 degree field of view and operations, including the data analysis,
and it will be fully automated.  The first data is expected to begin
to flow during 2001 and the total data volume will be approximately
four Terabytes. The limiting flux of the next stage of the ROTSE
experiment will be approximately $18$ -- $19$, or more than ten
times deeper than the original experiment. The ROTSE (and other
variability survey) data will provide important multi-epoch
measurements as a complement to the single epoch surveys (\eg DPOSS,
USNOA2, and the SDSS). Other examples of similar programs include the
Livermore Optical Transient Imaging System (LOTIS)
program\footnote{http://hubcap.clemson.edu/\twiddle ggwilli/LOTIS/},
which is nearly identical to the original ROTSE experiment, and its
successor, Super-LOTIS.

\item[NEAT] Near Earth Asteroid Tracking (NEAT) 
program\footnote{http://neat.jpl.nasa.gov/} is one of several programs
that are designed to discover and characterize near earth objects (\eg
Asteroids and comets). Fundamentally, these surveys cover thousands of
square degrees of the sky every month to a limiting flux depth of
approximately $17$ -- $20$, depending on the survey. All together,
these programs (which also include
Catalina\footnote{http://www.lpl.arizona.edu/css/},
LINEAR\footnote{http://www.ll.mit.edu/LINEAR/},
LONEOS\footnote{http://asteroid.lowell.edu/asteroid/loneos/loneos\_disc.html},
and Spacewatch\footnote{http://www.lpl.arizona.edu/spacewatch/})
generate nearly 200 Gigabytes of data a night, yet due to funding
restrictions, a large part of this data is not archived. NEAT is
currently the only program which provides archival access to its data
through the skymorph
project\footnote{http://skys.gsfc.nasa.gov/skymorph/skymorph.html}
(see Figure~\ref{neat} for the skymorph web-site). All told, these
surveys have around 10 Terabytes of imaging data in hand, and continue
to operate. The NEAT program has recently taken over the Palomar
Oschin 48-inch telescope, which was used to generate the two POSS
photographic surveys, in order to probe wider regions of the sky to an
even deeper limiting flux value.

\begin{figure}[!hbt]
\plotfiddle{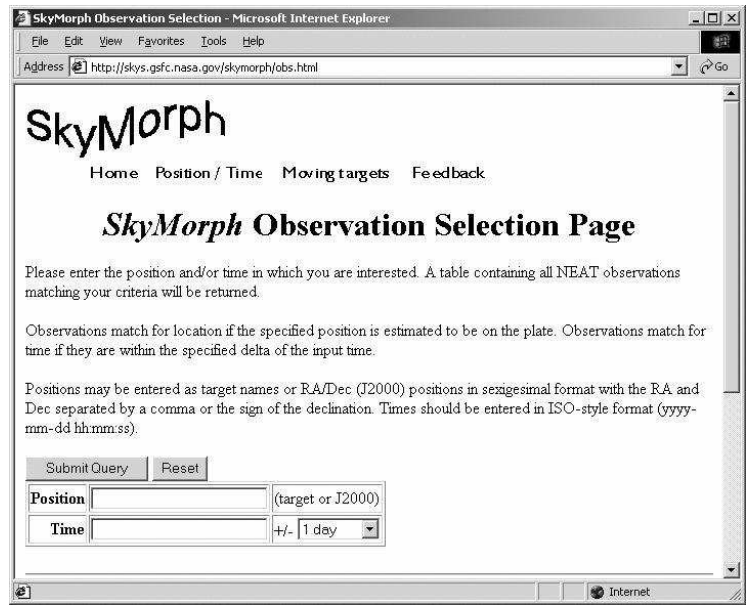}{3.0in}{0}{50}{50}{-140}{-9}
\caption{The skymorph web-site, which provides access to the data from the NEAT program.\label{neat}}
\end{figure}

\item[2MASS] The Two Micron All-Sky Survey\footnote{http://www.ipac.caltech.edu/2mass/} (2MASS) is a
near-infrared ($J$, $H$, and $K_S$) all sky survey. The 2MASS project
is a collaboration between the The University of Massachusetts, which
constructed the observatory facilities and operates the survey, and
the Infrared Processing and Analysis Center (Caltech), which is
responsible for all data processing and archive issues. The survey
began in the spring of 1997, completed survey quality observations in
2000, and is expected to publish the final catalog sometime in 2002.

The 2MASS survey utilized two new, highly automated 1.3m telescopes,
one at Mt. Hopkins, AZ and one at CTIO, Chile. Each telescope was
equipped with a three channel camera, which uses HgCdTe detectors, and
was capable of observing the sky simultaneously at $J$,$H$, and
$K_S$. The survey includes over twelve Terabytes of imaging data,
and the final catalog is expected to contain more than one million
resolved galaxies, and more than three hundred million stars and other
unresolved sources to a $10 \sigma$ limiting magnitude of $K_S < 14.3$.

The 2MASS program is leading the way in demonstrating the power of
public archives to the astronomical community. All twelve Terabytes of
imaging data are available on near-online tape storage, and the actual
catalogs are stored in an Informix backed archive (see
Figure~\ref{irsa} for the public access to the 2MASS data). When
complete, the survey will produce the following data products:

\begin{itemize}

\item a digital atlas of the sky comprising more than 1 million 8'x16' images having 
about 4\arcsec\ spatial resolution in each of the three wavelength
bands,

\item a point source catalog containing accurate (better than 0.5\arcsec) 
positions and fluxes (less than 5\% for $K_S > 13$) for approximately $300,000,000$ stars.

\item an extended source catalog containing positions and total magnitudes 
for more than 500,000 galaxies and other nebulae.

\end{itemize}

\item[NVSS] The National Radio Astronomical Observatory (NRAO), Very Large Array (VLA),
Sky Survey (NVSS) is a publicly available, radio continuum
survey\footnote{http://www.cv.nrao.edu/\twiddle jcondon/nvss.html} covering
the sky north of $-40$ degrees declination. The survey catalog
contains over 1.8 millions discrete sources with total intensity and
linear polarization image measurements (Stokes I, Q, and U) with a
resolution of 45\arcsec, and a completeness limit of about 2.5
mJy. The NVSS survey is now complete, containing over two hundred
thousand snapshot fields.The NVSS survey was performed as a community
service, with the principal data products being released to the public
by anonymous FTP as soon as they were produced and verified. The
primary means of accessing the NVSS data products remains the original
anonymous FTP service, however, other archive sites have begun to
provide limited data browsing and access to this important dataset.

The principal NVSS data products are a set of 2326 continuum map
``cubes,'' each covering an area of 4 degrees by 4 degrees with three
planes containing the Stokes I, Q, and U images, and a catalog of
discrete sources on these images (over 1.8 million sources are in the
entire survey).  Every large image was constructed from more than one
hundred of the smaller, original snapshot images.

\item[FIRST] The Faint Images of the Radio Sky at Twenty-cm (FIRST) 
survey\footnote{http://sundog.stsci.edu/top.html} is an ongoing,
publicly available, radio snapshot survey that is scheduled to cover
approximately ten thousand square degrees of the North and South
Galactic Caps in 1.8\arcsec\ pixels (currently, approximately eight
thousand square degrees have been released). The survey catalog, when
complete should contain around one million sources with a resolution
of better than 1\arcsec. The FIRST survey is being performed at the
NRAO VLA facility in a configuration that provides higher spatial
resolution than the NVSS, at the expense of a smaller field of view.

A final atlas of radio image maps with $1.8$\arcsec\ pixels is
produced by coadding the twelve images adjacent to each pointing
center. A source catalog including peak and integrated flux densities
and sizes derived from fitting a two-dimensional Gaussian to each
source is generated from the atlas. Approximately 15\% of the sources
have optical counterparts at the limit of the POSS I plates ($E \sim
20$); unambiguous optical identifications ($< 5$\% false rates) are
achievable to $V \sim 24$. The survey area has been chosen to coincide
with that of the Sloan Digital Sky Survey (SDSS). It is expected that
at the magnitude limit of the SDSS, approximately $50$\% of the
optical counterparts to FIRST sources will be detected. Both the
images and the catalogs constructed from the FIRST observations are
being made available to the astronomical community via the project
web-site (see Figure~\ref{first} for the public web-site) as soon as
sufficient quality-control tests have been completed.

\begin{figure}[!hbt]
\plotfiddle{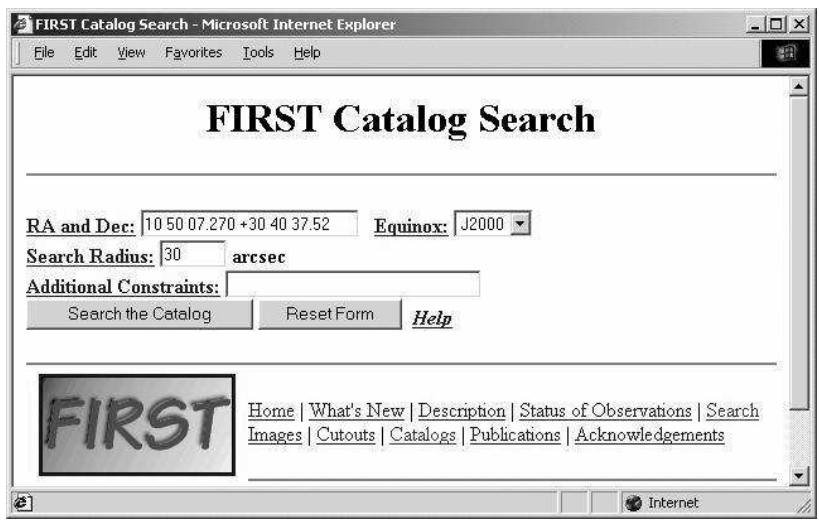}{2.75in}{0}{55}{55}{-160}{0}
\caption{The FIRST survey web search engine. Both images and catalog entries can be retrieved.\label{first}}
\end{figure}
\end{description}

\subsection{Ground-Based Observatory Archives}

Traditional ground-based observatories have been saving data, mainly
as emergency back-ups for the users, for a significant time,
accumulating impressive quantities of highly valuable, but
heterogeneous data. Unfortunately, with some notable exceptions, the
heterogeneity and a lack of adequate funding have limited the efforts
to properly archive this wealth of information and make it easily
available to the broad astronomical community. We see the development
of good archives for major ground-based observatories as one of the
most pressing needs in this field, and a necessary step in the
development of the National Virtual Observatory.

In this section, we discuss three specific ground-based observatories:
the National Optical Astronomical Observatories (NOAO), the National
Radio Astronomical Observatory (NRAO), and the European Southern
Observatory (ESO), focusing on their archival efforts. In addition to
these three, many other observatories have extensive archives,
including the Canada-France-Hawaii
telescope\footnote{http://www.cfht.hawaii.edu/} (CHFT), the James
Clerk Maxwell
Telescope\footnote{http://www.jach.hawaii.edu/JACpublic/JCMT/home.html}
(JCMT), the Isaac Newton Group of telescopes at La
Palma\footnote{http://www.ing.iac.es/} (ING), the Anglo-Australian
Observatory\footnote{http://site.aao.gov.au/AAO/} (AAT), the United
Kingdom Infrared
Telescope\footnote{http://www.jach.hawaii.edu/JACpublic/UKIRT/home.html}
(UKIRT), and the Australia National Telescope
Facility\footnote{http://www.atnf.csiro.au/} (ATNF).

\begin{description}

\item[NOAO] The National Optical Astronomy 
Observatory\footnote{http://www.noao.edu/} (NOAO) is a US organization
that manages ground-based, national astronomical observatories,
including the Kitt Peak National Observatory, Cerro Tololo
Inter-American Observatory, and the National Solar Observatory. NOAO
also represents the US astronomical community in the International
Gemini Project.  As a national facility, NOAO telescopes are open to
all astronomers regardless of institutional affiliation, and has
provided important scientific opportunities to astronomers throughout
the world, who would otherwise had little or no opportunity to obtain
astronomical observations.

The NOAO has been archiving all data from their telescopes in a
program called save-the-bits, which, prior to the introduction of
survey-grade instrumentation, generated around half a Terabyte and
over 250,000 images a year. With the introduction of survey
instruments and related programs, the rate of data accumulation has
increased, and NOAO now manages over 10 Terabytes of data. 

\item[NRAO] The National Radio Astronomy 
Observatory\footnote{http://www.nrao.edu/} (NRAO) is a US research
facility that provides access to radio telescope facilities for use by
the scientific community, in analogy to the primarily optical mission
of the NOAO. Founded in 1956, the NRAO has its headquarters in
Charlottesville, VA, and operates major radio telescope facilities at
Green Bank, WV, Socorro, NM, and Tucson, AZ. The NRAO has been
archiving their routine observations and has accumulated over ten
Terabytes of data. They also have provided numerous opportunities for
surveys, including the previously discussed NVSS and FIRST radio
surveys as well as the Green Bank surveys.

\item[ESO] The European Southern Observatory\footnote{http://www.eso.org/}
(ESO), operates a number of telescopes (including the four 8m class
VLT telescopes) at two observatories in the southern hemisphere: the
La Silla Observatory, and the Paranal observatory. ESO is currently
supported by a consortium of countries, with Headquarters in Garching,
near Munich, Germany.

As with many of the other ground-based observatories, ESO has been
archiving data for some time, with two important differences. First,
they were one of the earliest observatories to appreciate the
importance of community service survey programs (these programs
generally probe to fainter flux limits over a significantly smaller
area than the previously discussed surveys), which are made accessible
to the international astronomical community in a relatively short
timescale. Second, appreciating the legacy aspects of the four 8 meter
Very Large Telescopes, ESO intentionally decided to break with
tradition, and imposed an automatic operation of the telescopes that
provides a uniform mechanism for data acquisition and archiving,
comparable to what has routinely been done for space-based
observatories (see the next section). Currently, the ESO data archive
is starting to approach a steady state rate of approximately 20
Terabytes of data per year from all of their telescopes. This number
will eventually increase to several hundred Terabytes with the
completion of the rest of the planned facilities, including the VST, a
dedicated survey telescope similar in nature to the telescope that was
built for the SDSS project.

\end{description}

\subsection{NASA Space-Based Observatory Archives}

With the continual advancement of technology, ground-based
observations continue to make important discoveries. Our atmosphere,
however, absorbs radiation from the majority of the electromagnetic
spectrum, which, while important to the survival of life, is a major
hindrance when trying to untangle the mysteries of the cosmos. Thus
space-based observations are critical, yet they are extremely
expensive. The resulting data is extremely valuable, and all of the
generated data is archived. While there have been (and continue to be)
a large number of satellite missions, we will focus on three major
NASA archival centers: MAST, IRSA, and HEASARC (officially designated
as NASA's distributed Space Science Data Services), the Chandra X-ray
Observatory archive (CXO), and the National Space Science Data Center
(NSSDC).

\begin{description}

\item[MAST] The Multimission Archive at the Space Telescope Science 
Institute\footnote{http://archive.stsci.edu/mast.html} (MAST) archives
a variety of astronomical data, with the primary emphasis on the
optical, ultraviolet, and near-infrared parts of the spectrum. MAST
provides a cross correlation tool that allows users to search all
archived data for all observations which contain sources from either
archived or user supplied catalog data. In addition, MAST provides
individual mission query capabilities. Preview images or spectra can
often be obtained, which provides useful feedback to archive
users. The dominant holding for MAST is the data archive from the
Hubble Space Telescope (see Figure~\ref{mast} for the HST archive
web-site). This archive has been replicated at mirror sites in
Canada\footnote{http://cadcwww.dao.nrc.ca/hst/},
Europe\footnote{http://www.stecf.org/} and
Japan\footnote{http://hst.nao.ac.jp/}, and has often taken a lead in
astronomical archive developments.

\begin{figure}[!hbt]
\plotfiddle{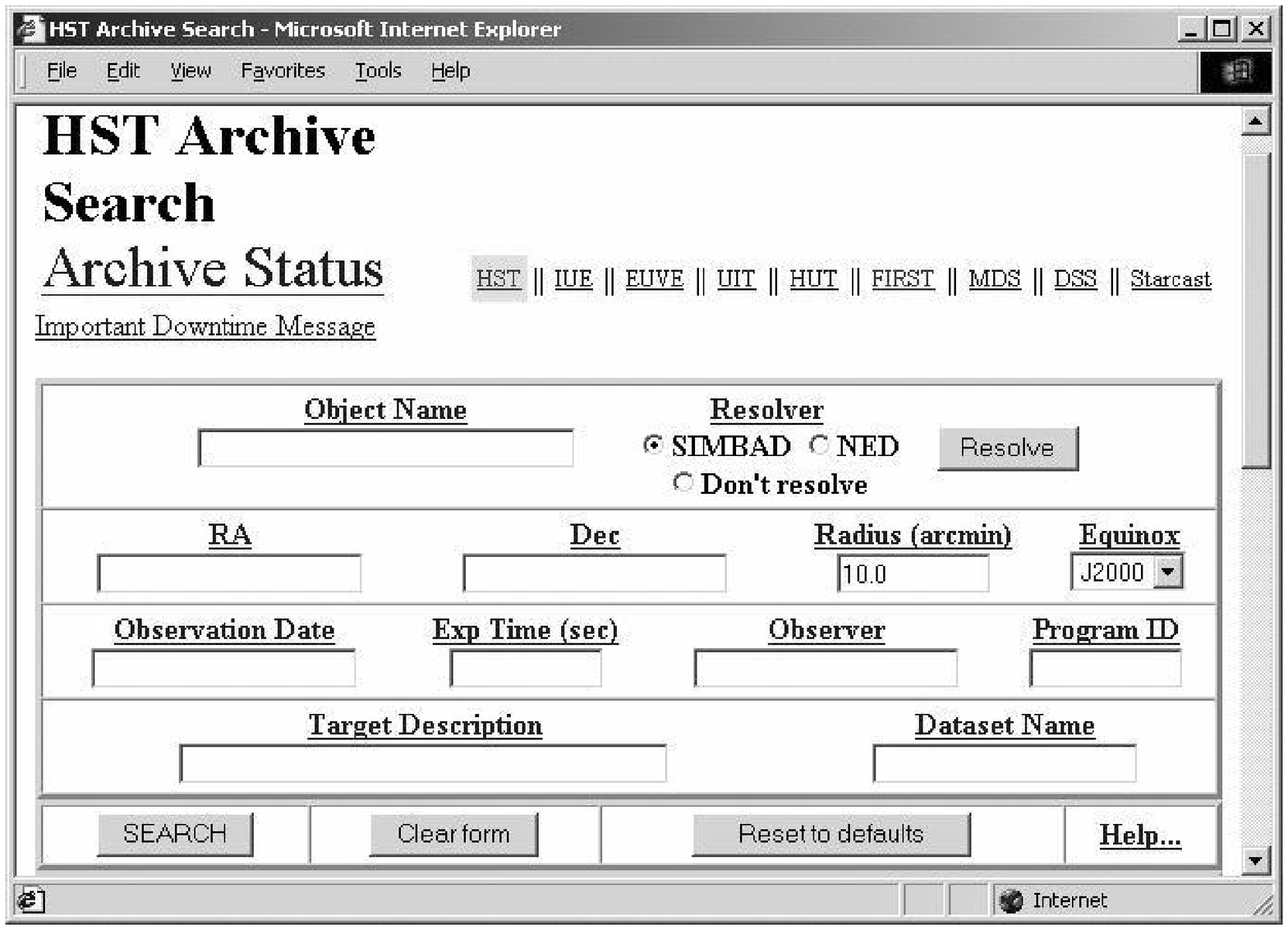}{3.25in}{0}{55}{55}{-160}{-9}
\caption{The Hubble Space Telescope (HST) archive (part of the MAST holdings) interface.\label{mast}}
\end{figure}

Based on the archival nature of the requested data, MAST provides data
access in a variety of different ways, including intermediate disk
staging and FTP retrieval and direct web-based downloads. MAST holdings
currently exceed ten Terabytes, including or providing links to
archival data for the following missions or projects:

\begin{itemize}

\item the Hubble Space Telescope (HST),
\item the International Ultraviolet Explorer (IUE),
\item Copernicus UV Satellite (OAO-3),
\item the Extreme Ultraviolet Explorer (EUVE),
\item the two space shuttle ASTRO Missions:
\begin{itemize}
\item the Ultraviolet Imaging Telescope (UIT),
\item the Wisconsin Ultraviolet Photo Polarimetry Experiment (WUPPE),
\item the Hopkins Ultraviolet Telescope (HUT),
\end{itemize}
\item the FIRST (VLA radio data) survey,
\item the Digitized Sky Survey,
\item the Roentgen Satellite (ROSAT) archive,
\item the Far Ultraviolet Spectroscopic Explorer (FUSE),
\item Orbiting and Retrieval Far and Extreme Ultraviolet Spectrograph (ORFEUS) missions,
\item the Interstellar Medium Absorption Profile Spectrograph(IMAPS) (first flight),
\item the Berkeley Extreme and Far-UV Spectrometer (BEFS) (first flight).
\end{itemize}

The Hubble Space Telescope (HST) is the first of NASA's great
observatories, and provides high-resolution imaging and spectrographic
observations from the near-ultraviolet to the near-infrared parts of
the electro-magnetic spectrum ($0.1$ -- $2.5$ microns). It is operated
by the Space Telescope Science Institute (STScI) is located on the
campus of the Johns Hopkins University and is operated for NASA by the
Association of Universities for Research in Astronomy (AURA).

\item[IRSA] The NASA Infrared Processing and Analysis Center (IPAC) Infrared Science 
Archive\footnote{http://irsa.ipac.caltech.edu} (IRSA) provides
archival access to a variety of data, with a primary focus on data in
the infrared portion of the spectrum (see Figure~\ref{irsa} for the
IRSA archive web-site). IRSA has taken a strong leadership position in
developing software and Internet services to facilitate access to
astronomical data products. IRSA provides a cross-correlation tool
allowing users to extract specific data on candidate targets from a
variety of sources. IRSA provides primarily browser based query
mechanisms, including access to both catalog and image holdings.

\begin{figure}[!hbt]
\plotfiddle{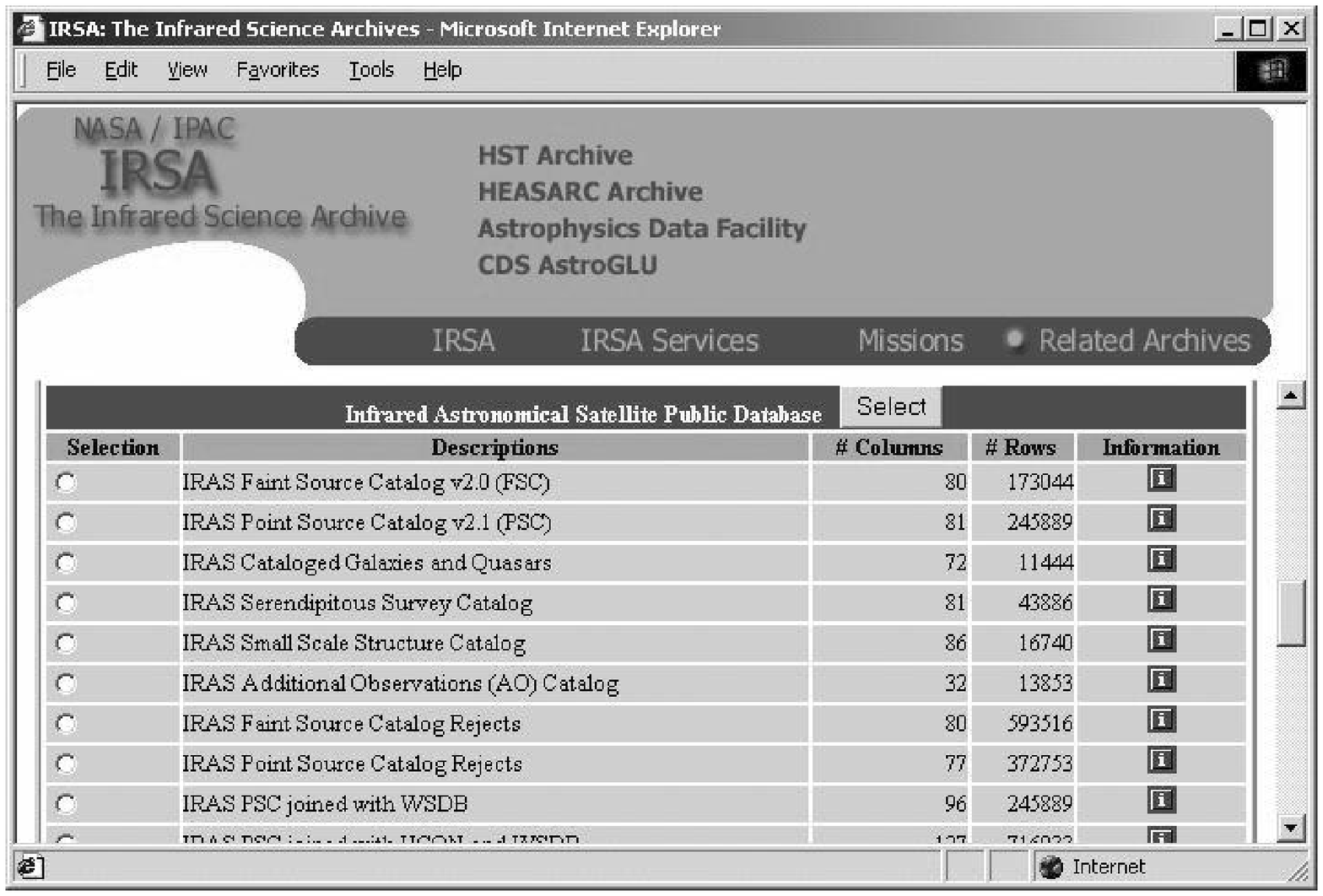}{3.1in}{0}{55}{55}{-160}{-9}
\caption{The IRSA Public access site. From this web-site, an archive user 
can query the contents of the publicly available 2MASS data, as well
as other datasets that have been ingested into the IRSA Informix
database server.\label{irsa}}
\end{figure}

IRSA contains over fifteen Terabytes of data, mostly related to the
2MASS survey, and currently maintains the archives for the following
datasets at IPAC:

\begin{itemize}
\item the Two Micron All-Sky Survey (2MASS),
\item the Space Infrared Telescope Facility (SIRTF),
\item the European Infrared Space Observatory (ISO),
\item the Midcourse Space Experiment (MSX),
\item the Infrared Astronomical Satellite (IRAS).
\end{itemize}

The Space Infrared Telescope (SIRTF) observatory will be the fourth
and final of NASA's great observatories, and will provide imaging and
spectrographic observations in the infrared part of the
electro-magnetic spectrum ($3$ -- $180$ microns). SIRTF is expected to
be launched in 2002. The SIRTF science center (SSC) is located on the
campus of the California Institute of Technology and is operated for
NASA by the Jet propulsion Laboratory.

The Infrared Space Observatory (ISO) operated at wavelengths from
$2.5$ -- $240$ microns, obtaining both imaging and spectroscopic data
are available over a large area of the sky. The Midcourse Space
Experiment (MSX) operated from $4.2$ -- $26$ microns, and mapped the
Galactic Plane, the gaps in the IRAS data, the zodiacal background,
confused regions away from the Galactic Plane, deep surveys of
selected fields at high galactic latitudes, large galaxies, asteroids
and comets.  The Infrared Astronomical Satellite (IRAS) performed an
unbiased all sky survey at 12, 25, 60 and 100 microns, detecting
approximately $350,000$ high signal-to-noise infrared sources split
between the faint and point source catalogs. A significantly larger
number of sources ($> 500,000$) are included in the faint source
reject file, which were below the flux threshold required for the
faint source catalog.

\item[HEASARC] The High Energy Astrophysics Science Archive
Research Center\footnote{http://heasarc.gsfc.nasa.gov} (HEASARC) is a
multi-mission astronomy archive with primary emphasis on the extreme
ultra-violet, X-ray, and Gamma ray spectral regions.  HEASARC
currently holds data from 20 observatories covering 30 years of X-ray
and gamma-ray astronomy. HEASARC provides data access via FTP and the
Web (see Figure~\ref{heasarc}), including the Skyview interface, which
allows multiple images and catalogs to be compared.

\begin{figure}[!hbt]
\plotfiddle{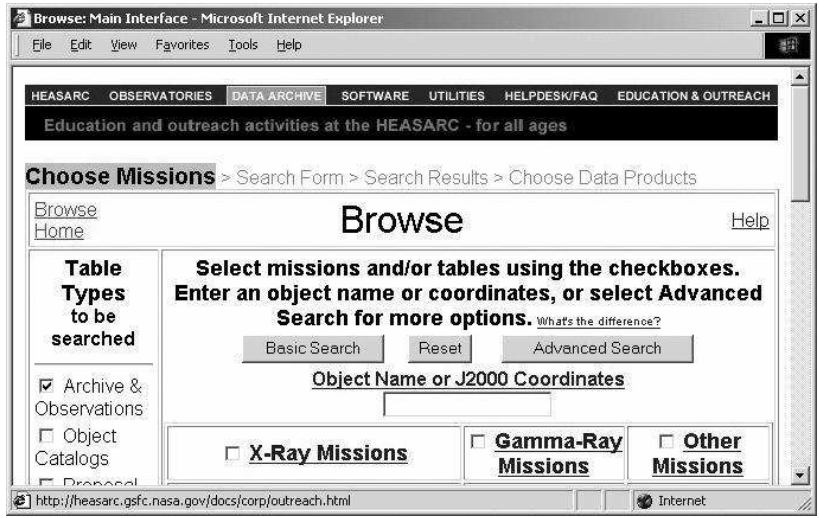}{3.0in}{0}{55}{55}{-160}{-9}
\caption{The HEASARC public access site, where data from the x-ray and 
gamma-ray space missions are archived.\label{heasarc}}
\end{figure}

The HEASARC archive currently includes over five Terabytes of data,
and will experience significant increases with the large number of
high-energy satellite missions that are either currently or soon will
be in operation. HEASARC currently provides archival access to the
following missions:

\begin{itemize}
\item the Roentgen Satellite (ROSAT),
\item the Advanced Satellite for Cosmology (ASCA),
\item BeppoSAX, 
\item the Compton Gamma Ray Observatory (CGRO), the second of NASA's great observatories, 
\item the Extreme Ultraviolet Explorer (EUVE), 
\item the High Energy Astrophysics Observatory (HEAO 1), 
\item the Einstein Observatory (HEAO 2), 
\item the European Space Agency's X-ray Observatory (EXOSAT), 
\item the Rossi X-ray Timing Explorer (Rossi XTE), 
\item the Advanced Satellite for X-ray Astronomy (Chandra), 
\item the X-ray Multiple Mirror Satellite (XMM-Newton), 
\item the High Energy Transient Explorer (HETE-2), 
\end{itemize}

Several of these missions elicit further discussion. First, the
Einstein
Observatory\footnote{http://heasarc.gsfc.nasa.gov/docs/einstein/heao2.html}
was the first fully imaging X-ray telescope put into space, with an
angular resolution of a few arcseconds and was sensitive over the
energy range $0.2$ -- $3.5$ keV. The
ROSAT\footnote{http://wave.xray.mpe.mpg.de/rosat/} satellite was an
X-ray observatory that performed an all sky survey in the $0.1$ --
$2.4$ keV range as well as numerous pointed observations. The full sky
survey data has been publicly released, and there are catalogs of both
the full sky survey~\citep{voges99} as well as serendipitous
detections from the pointed observations~\citep{white94}. The
ASCA\footnote{http://www.astro.isas.ac.jp/asca/index-e.html} satellite
operated in the energy range $0.4$ -- $10$ keV, that performed several
small area surveys and obtained valuable spectral data on a variety of
astrophysical sources. Finally, the Chandra (also see the next
section) and XMM-Newton\footnote{http://xmm.vilspa.esa.es/} satellites
are currently providing revolutionary new views on the cosmos, due to
their increased sensitivity, spatial resolution and collecting area.

\item[Skyview\footnote{http://skyview.gsfc.nasa.gov/}] is a web-site 
operated from HEASARC which is billed as a ``Virtual Observatory on
the Net'' (see Figure~\ref{skyview}). Using public-domain data,
Skyview allows astronomers to generate images of any portion of the
sky at wavelengths in all regimes from radio to gamma-ray. Perhaps the
most powerful feature of the Skyview site is its ability to handle the
geometric and coordinate transformations required for presenting the
requested data to the user in the specified format.

\begin{figure}[!hbt]
\plotfiddle{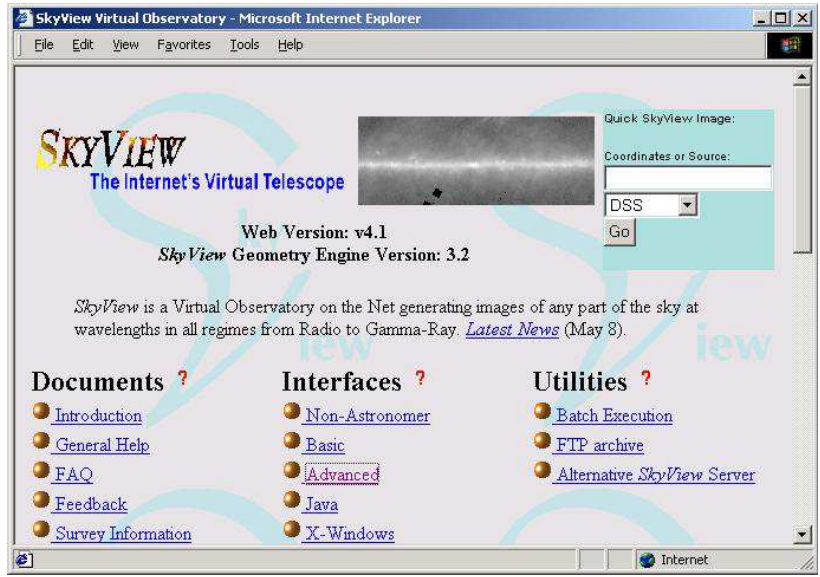}{3.15in}{0}{55}{55}{-165}{-9}
\caption{The Skyview virtual telescope site, which provides access to a 
large number of digital imagery and catalog
information.\label{skyview}}
\end{figure}

\item[CDA] The Chandra X-ray observatory\footnote{http://chandra.harvard.edu/} 
is third of NASA's great observatories, and provides high-resolution
imaging and spectrographic observations in the X-ray part of the
electro-magnetic spectrum. Chandra was launched by the Space Shuttle
Columbia during July, 1999. Unlike the Hubble data archive, which is
part of MAST, and the SIRTF data archive, which will be part of IRSA,
the Chandra Data Archive (CDA) is part of the Chandra X-Ray
Observatory Science Center (CXC) which is operated for NASA by the
Smithsonian Astrophysical Observatory. The data is also archived at
HEASARC, which is the relevant wavelength space mission archive.

The Chandra data products can be roughly divided into science-related
and engineering data. The engineering data products include all data
relating to the spacecraft subsystems: including such quantities as
spacecraft temperature and operating voltages. The scientific data
products are divided into three categories: primary, secondary, and
supporting products. Primary products are generally the most desired,
but the secondary products can provide important information required
for more sophisticated analysis and, possibly, limited reprocessing
and fine-tuning. The actual data can be retrieved via several
different mechanisms or media, including web-based downloads, staged
anonymous FTP retrieval, or mailed delivery of 8 mm Exabyte, 4 mm DAT,
or CDROMs.

\item[NSSDC] The National Space Science Data 
Center\footnote{http://nssdc.gsfc.nasa.gov/} (NSSDC) provides
network-based and offline access to a wide variety of data from NASA
missions, including the Cosmic Background
Explorer\footnote{http://space.gsfc.nasa.gov/astro.cobe} (COBE),
accruing data at the rate of several Terabytes per year. NSSDC was
first established at the Goddard Space Flight Center in 1966, and
continues to archive mission data, including both independently and
not independently (\ie data that requires other data to gain utility)
useful data. Currently, the majority of network-based data
dissemination is via WWW and FTP, and most offline data dissemination
is via CD-ROM. NSSDC is generally regarded as the final repository of
all NASA space mission data.

\item[PDS] In this chapter, we do not discuss in detail the great 
wealth of data available on solar system objects. The site for the
archival access to scientific data from NASA planetary missions,
astronomical observations, and laboratory measurements is the
Planetary Data System\footnote{http://pds.jpl.nasa.gov} (PDS). The
homepage for the PDS is displayed in Figure~\ref{pds}, and provides
access to data from the Pioneer, Voyager, Mariner, Magellan, NEAR
spacecraft missions, as well as other data on asteroids, comets, and
the Planets---including Earth.

\begin{figure}[!hbt]
\plotfiddle{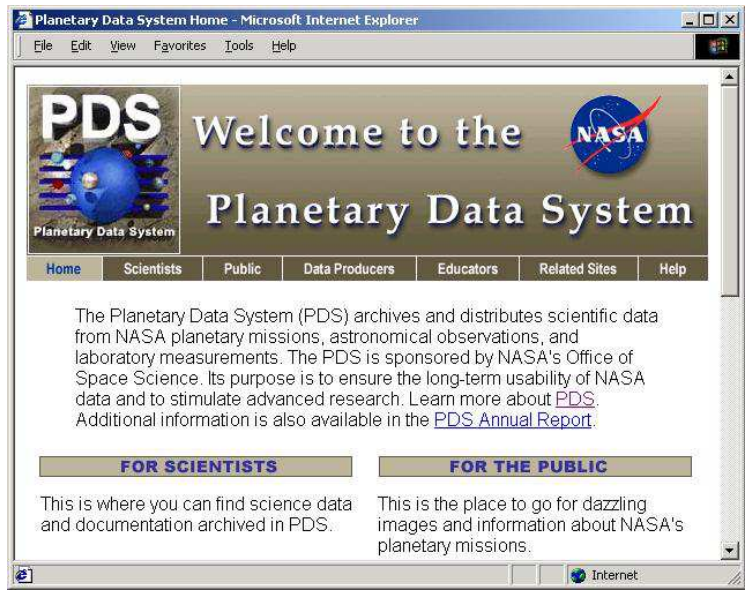}{3.25in}{0}{50}{50}{-155}{0}
\caption{The Planetary Data System web-site. From this web-site, 
users can query and extract archived the enormous quantity of data
that has been obtained on the Solar system objects.\label{pds}}
\end{figure}
\end{description}

\section{The Future of Observational Astronomy: Virtual Observatories\label{vo-initaive}}

Raw data, no matter how expensively obtained, are no good without an
effective ability to process them quickly and thoroughly, and to
refine the essence of scientific knowledge from them.  This problem
has suddenly increased by orders of magnitude, and it keeps growing.

A prime example is the efficient scientific exploration of the new
multi-Terabyte digital sky surveys and archives.  How can one make
efficiently discoveries in a database of billions of objects or data
vectors?  What good are the vast new data sets if we cannot fully
exploit them?  

In order to cope with this data flood, the astronomical community
started a grassroots initiative, the National (and ultimately Global)
Virtual Observatory~\citep[see, \eg][for a virtual observatory
conference proceedings]{NVO2K}.  Recognizing the urgent need, the
National Academy of Science Astronomy and Astrophysics Survey
Committee, in its new decadal survey entitled {\em Astronomy and
Astrophysics in the New Millennium}, recommends, as a first priority,
the establishment of a National Virtual Observatory. The NVO will
likely grow into a Global Virtual Observatory, serving as the
fundamental information infrastructure for astronomy and astrophysics
in the next century.  We envision productive international cooperation
in this rapidly developing new field.

The NVO would federate numerous large digital sky archives, provide
the information infrastructure and standards for ingestion of new data
and surveys, and develop the computational and analysis tools with
which to explore these vast data volumes.  It would provide new
opportunities for scientific discovery that were unimaginable just a
few years ago.  Entirely new and unexpected scientific results of
major significance will emerge from the combined use of the resulting
datasets, science that would not be possible from such sets used
singly.  The NVO will serve as {\em an engine of discovery for
astronomy}~\citep{nvo-white}.

Implementation of the NVO involves significant technical challenges on
many fronts: How to manage, combine, analyze and explore these vast
amounts of information, and to do it quickly and efficiently?  We know
how to collect many bits of information, but can we effectively refine
the essence of knowledge from this mass of bits?  Many individual
digital sky survey archives, servers, and digital libraries already
exist, and represent essential tools of modern astronomy.  However, in
order to join or federate these valuable resources, and to enable a
smooth inclusion of even greater data sets to come, a more powerful
infrastructure and a set of tools are needed. 

The rest of this review focuses on the two core challenges that must
be tackled to enable the new, virtual astronomy:

\begin{enumerate}

\item
Effective federation of large, geographically distributed data sets
and digital sky archives, their matching, their structuring in new
ways so as to optimize the use of data-mining algorithms, and fast
data extraction from them.

\item 
Data mining and ``knowledge discovery in databases'' (KDD) algorithms
and techniques for the exploration and scientific utilization of large
digital sky surveys, including combined, multi-wavelength data sets.

\end{enumerate}

These services would carry significant relevance beyond Astronomy as
many aspects of society are struggling with information overload. This
development can only be done by a wide collaboration, that involves
not only astronomers, but computer scientists, statisticians and even
participants from the IT industry.

\subsection{Architecting the Virtual Observatory}

The foundation and structure of the National Virtual Observatory (NVO)
are not yet clearly defined, and are currently the subject of many
vigorous development efforts.  One framework for many of the basic
architectural concepts and associated components of a virtual
observatory, however, has become popular. First is the requirement
that, if at all possible, all data must be maintained and curated by
the respective groups who know it best --- the survey
originators. This requires a fully distributed system, as each survey
must provide the storage, documentation, and services that are
required to participate in a virtual observatory.

The interconnection of the different archive sites will need to
utilize the planned high-speed networks, of which there are several
testbed programs already available or in development.  A significant
fraction of the technology for the future Internet backbone is already
available, the problem is finding real-world applications which can
provide a sufficient load. A Virtual Observatory, would, of course,
provide heavy network traffic and is, therefore, a prime candidate for
early testing of any future high-speed networks.

\begin{figure}[!hbt]
\plotfiddle{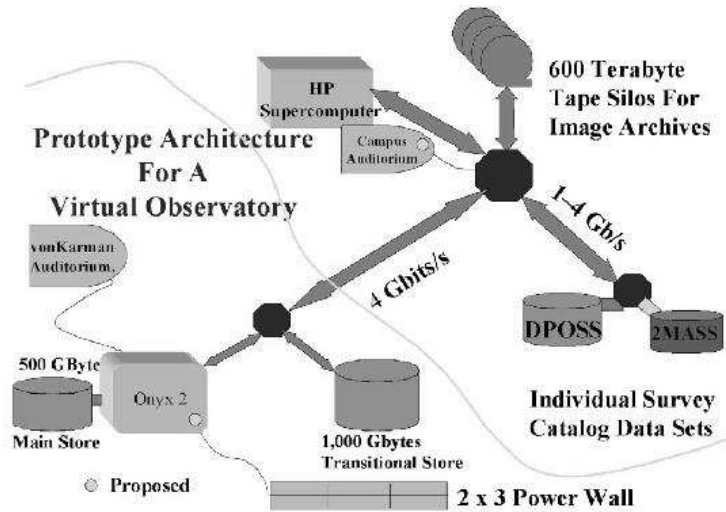}{3.0in}{0}{50}{50}{-150}{0}
\caption{One straw-man layout for a virtual observatory prototype, demonstrating the 
highly distributed nature of the system. Of particular importance are
the high speed network interconnections, extremely large storage
systems, and high performance compute servers. Image Credit: Dave
Curkendall and the ALPHA group, (JPL).\label{cit-vo}}
\end{figure}

The distributed approach advocated by this framework (see
Figure~\ref{cit-vo} for a demonstration) relies heavily on an the
ability of different archives to participate in ``collaborative
querying''. This tight integration requires that everything must be
built using appropriately developed standards, detailing everything
from how archives are ``discovered'' and ``join'' the virtual
observatory, to how queries are expressed and data is
transferred. Once these standards have been developed, implementation
(or retrofitting as the case may be) of tools, interfaces, and
protocols that operate within the virtual observatory can begin. 

The architecture of a virtual observatory is not only dependent on the
participating data centers, but also on the users it must support. For
example, it is quite likely that the general astronomy public (\eg
amateur astronomers, K-12 classrooms, {\em etc.}) would use a virtual
observatory in a casual lookup manner (\ie the web model). On the
other hand, a typical researcher would require more complex services,
such as large data retrieval (\eg images) or cross-archive
joins. Finally, there will also be the ``power users'' who would
require heavy post-processing of query results using super-computing
resources (\eg clustering analysis).

From these user models we can derive ``use cases'', which detail how
a virtual observatory might be utilized. Initially, one would expect a
large number of distinct ``exploratory'' queries as astronomers
explore the multi-dimensional nature of the data. Eventually the
queries will become more complex and employ a larger scope or more
powerful services. This model requires the support of several methods
for data interaction: manual browsing, where a researcher explores the
properties of an interesting class of objects; cross-identification
queries, where a user wants to find all known information for a given
set of sources; sweeping queries, where large amounts of data (\eg
large areal extents, rare object searches) are processed with complex
relationships; and the creation of new ``personal'' subsets or
``official'' data products. This approach leads, by necessity, to
allowing the user to perform customizable, complex analysis (\ie
data-mining) on the extracted data stream.

\subsection{Connecting Distributed Archives}

\begin{figure}[!hbt]
\plotfiddle{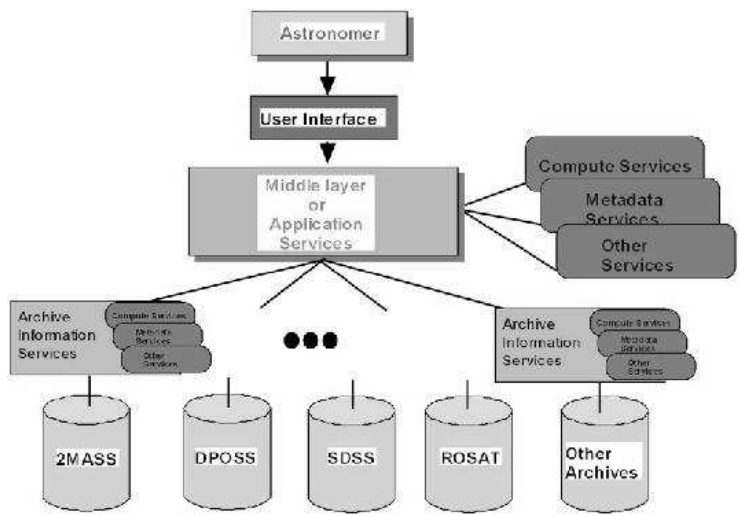}{3.0in}{0}{50}{50}{-150}{0}
\caption{A prototype blueprint for the system architecture of a virtual observatory. 
The key concept throughout our approach is the plug---and---play model
where different archives, compute services and user tools all interact
seamlessly. This system model is predicated on the universal adoption
of standards dictating everything from how archives communicate with
each other to how data is transferred between archives, services and
users.\label{as}}
\end{figure}

As can be seen from Section~\ref{dataset}, a considerable amount of
effort has been expended within the astronomical community on
archiving and processing astronomical data. On the other hand, very
little has been accomplished in enabling collaborative, cross-archive
data manipulation (see Figure~\ref{as}). This has been due, in part,
to the previous dearth of large, homogeneous, multi-wavelength
surveys; in other words, the payoff for federating the disparate
datasets has previously been too small to make the effort
worthwhile. Here we briefly outline some of the key problem
areas~\citep[\cf][]{fgcs} for a more detailed discussion), that must
be addressed in order to properly build the foundation for the future
virtual observatories.

\subsubsection{Communication Fundamentals}

The first requirement for connecting highly distributed datasets is
that they must be able to communicate with each other. This
communication takes multiple roles, including the initiation of
communication, discovering the holdings and capabilities of each
archive, the actual process of querying, the streaming of data, and an
overall control structure. None of these ideas are entirely new, the
general Information Technology field has been confronting similar
issues and solutions, such as Grid frameworks~\citep{grid},
JINI\footnote{A registered trademark of the SUN Microsystems
corporation.} and the Web services model (see, \eg the IBM Web Service
web-site\footnote{http://www.ibm.com/developerworks/webservices/}) for
more information) are equally applicable.

Clearly, the language for communicating will be the extensible markup
language (XML), using a community defined standard schema. This will
allow for control of the inter-archive communication and processing
(\eg the ability to perform basic checkpoint operations on a query:
stop, pause, restart, abort, and provide feedback to the end-user). A
promising, and simple mechanism for providing the archive
communication endpoints is through web services, which would be built
using the Simple Object Access
Protocol\footnote{http://www.w3.org/TR/2000/NOTE-SOAP-20000508/}
(SOAP), Web Services Description
Language\footnote{http://xml.coverpages.org/wsdl.html} (WSDL), and a
common Universal Description, Discovery, and
Integration\footnote{http://www.uddi.org/} (UDDI) registry.  An
additional benefit of this approach is that pre-existing or legacy
archival services can be retrofitted (by mapping a new service onto
existing services) in order to participate in collaborative querying.

This model also allows for certain optimizations to be performed
depending on the status of the archival connections (\eg network
weather).  Eventually, a learning mechanism can be applied to analyze
queries, and using the accumulated knowledge gained from past
observations (\ie artificial intelligence), queries can be rearranged
in order to provide further performance enhancements.

\subsubsection{Archival Metadata}

In order for the discovery process to be successful, the archives must
communicate using shared semantics. Not only must this semantic format
allow for the transfer of data contents and formats between archives,
but it also should clearly describe the specific services that an
archive can support (such as cross-identification or image
registration) and the expected format of the input and output
data. Using the web service model, our services would be registered in
a well known UDDI registry, and communicate their capabilities using
WSDL. Depending on the need of the consumer, different amounts (or
levels) of detailed information might be required, leading to the need
for a hierarchical representation. Once again, the combination of XML
and a standardized XML Schema language provides an extremely powerful
solution, as is easily generated, and can be parsed by machines and
read by humans with equal ease. By adopting a standardized schema,
metadata can be easily archived and accessed by any conforming
application.

\subsubsection{High Performance Data Streaming}

Traditionally, astronomers have communicated data either in ASCII text
(either straight or compressed), or by using the community standard
Flexible Image Transport Standard (FITS). The true efficacy of the
FITS format as a streaming format, however, is not clear, due to the
difficulty of randomly extracting desired data or shutting off the
stream. The ideal solution would pass different types of data (\ie
tabular, spectral, or imaging data) in a streaming fashion (similar to
MPI---Message Passing Interface), so that analysis of the data does
not need to wait for the entire dataset before proceeding. In the web
services model, this would allow different services to cooperate in a
head-to-tail fashion (\ie the UNIX pipe scenario). This is still a
potential concern, as the ability to handle XML encoded binary data
is not known.

\subsubsection{Astronomical Data Federation}

Separate from the concerns of the physical federation of astronomical
data via a virtual observatory paradigm is the issue of actually
correlating the catalog information from the diverse array of
multiwavelength data (see the skyserver
project~\footnote{http://www.skyserver.org/} for more
information). While seemingly simple, the problem is complicated by
the several factors.

First is the sheer size of the problem, as the cross-identification of
billions of sources in both a static and dynamic state over thousands
of square degrees in a multi-wavelength domain (Radio to X-Ray) is
clearly a computationally challenging problem, even for a consolidated
archive. The problem is further complicated by the fact that
observational data is always limited by the available technology,
which varies greatly in sensitivity and angular resolution as a
function of wavelength (\eg optical-infrared resolution is generally
superior to high energy resolution).

Furthermore, the quality of the data calibration (either
spectral, temporal, or spatial) can also vary greatly, making it
extremely difficult to to unambiguously match sources between
different wavelength surveys. Finally, the sky looks different at
different wavelengths (see, \eg Figure~\ref{crab}), which can produce
one-to-one, many-to-one, one-to-one, many-to-many, and even
one/many-to-none scenarios when federating multiwavelength
datasets. As a result, sometimes the source associations must be made
using probabilistic methods~\citep{lonsdale98,rutledge2k}.

\subsection{Data Mining and Knowledge Discovery}

\begin{figure}[!hbt]
\plotone{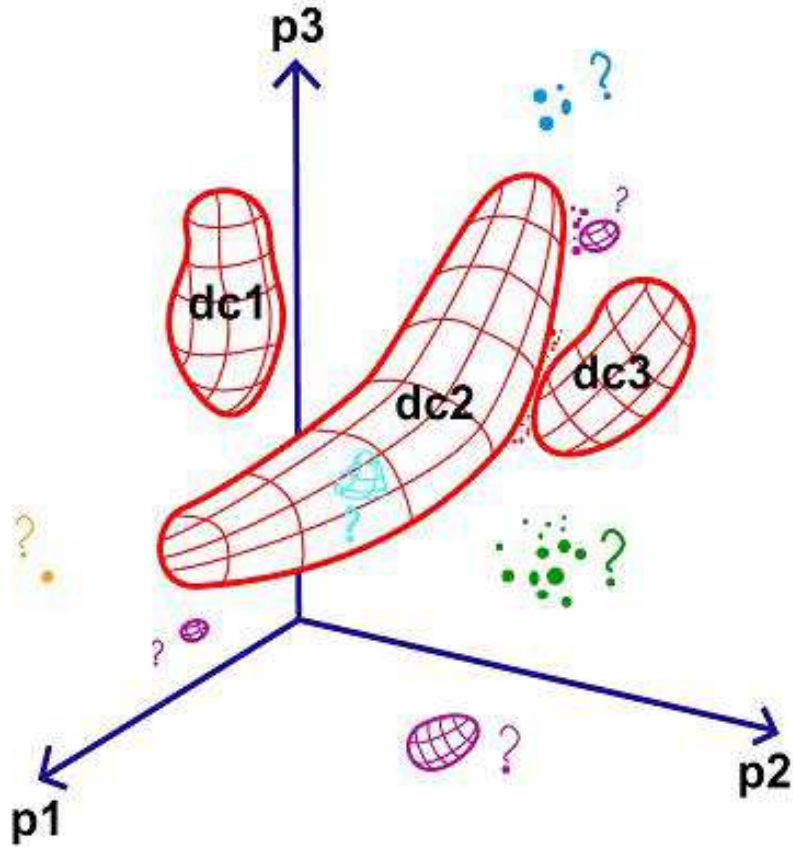}
\caption{A demonstration of a generic machine-assisted discovery problem---data 
mapping and a search for outliers. This schematic illustration is of
the clustering problem in a parameter space given by three object
attributes: P1, P2, and P3. In this example, most of the data points
are assumed to be contained in three, dominant clusters (DC1, DC2, and
DC3). However, one may want to discover less populated clusters (\eg
small groups or even isolated points), some of which may be too
sparsely populated, or lie too close to one of the major data
clouds. All of them present challenges of establishing statistical
significance, as well as establishing membership. In some cases,
negative clusters (holes), may exist in one of the major data
clusters.
\label{param-space}}
\end{figure}

Key to maximizing the knowledge extracted from the ever-growing
quantities of astronomical (or any other type of) data is the
successful application of data-mining and knowledge discovery
techniques. This effort as a step towards the development of the next
generation of science analysis tools that redefine the way scientists
interact and extract information from large data sets, here
specifically the large new digital sky survey archives, which are
driving the need for a virtual observatory (see, \eg
Figure~\ref{param-space} for an illustration).

Such techniques are rather general, and should find numerous
applications outside astronomy and space science. In fact, these
techniques can find application in virtually every data-intensive
field.  Here we briefly outline some of the applications of these
technologies on massive datasets, namely, unsupervised clustering,
other Bayesian inference and cluster analysis tools, as well as novel
multidimensional image and catalog visualization techniques. Examples
of particular studies may include:

\begin{enumerate}

\item 
Various classification techniques, including decision tree ensembles
and nearest-neighbor classifiers to categorize objects
or clusters of objects of interest.  Do the objectively found groupings
of data vectors correspond to physically meaningful, distinct types of
objects?  Are the known types recovered, and are there new ones?  Can we
refine astronomical classifications of object types (\eg the Hubble
sequence, the stellar spectral types) in an objective manner?

\item 
Clustering techniques, such as the expectation maximization (EM)
algorithm with mixture models to find groups of interest, to
come up with descriptive summaries, and to build density
estimates for large data sets.  How many distinct types of objects are
present in the data, in some statistical and objective sense?  This
would also be an effective way to group the data for specific studies,
\eg some users would want only stars, others only galaxies, or only
objects with an IR excess, \etc

\item
Use of genetic algorithms to devise improved detection and supervised
classification methods.  This would be especially interesting in the 
context of interaction between the image (pixel) and catalog (attribute)
domains.

\item 
Clustering techniques to detect rare, anomalous, or somehow unusual
objects, \eg as outliers in the parameter space, to be selected
for further investigation.  This would include both known but rare
classes of objects, \eg brown dwarfs, high-redshift quasars, and also
possibly new and previously unrecognized types of objects and phenomena.

\item 
Use of semi-autonomous AI or software agents to explore the large data
parameter spaces and report on the occurrences of unusual instances or
classes of objects.  How can the data be structured to allow for an
optimal exploration of the parameter spaces in this manner?

\item 
Effective new data visualization and presentation techniques, which can
convey most of the multidimensional information in a way more easily
grasped by a human user.  We could use three graphical dimensions, plus
object shapes and coloring to encode a variety of parameters, and to
cross-link the image (or pixel) and catalog domains.

\end{enumerate}

Notice that the above examples are moving beyond merely providing
assistance with handling of huge data sets: these software tools may
become capable of {\it independent or cooperative discoveries}, and
their application may greatly enhance the productivity of practicing
scientists.

It is quite likely that many of the advanced tools needed for these
tasks already exist or be under development in the various fields of
computer science and statistics.  In creating a virtual observatory,
one of the most important requirements is to bridge the gap between
the disciplines, and introduce modern data management and analysis
software technologies into astronomy and astrophysics.

\subsubsection{Applied Unsupervised Classification}

Some preliminary and illusory experiments using Bayesian clustering
algorithms were designed to classify objects present in the DPOSS
catalogs~\citep{decarvalho95} using the AutoClass
software~\citep{cheeseman88}.  The program was able separate the data
into four recognizable and astronomically meaningful classes: stars,
galaxies with bright central cores, galaxies without bright cores, and
stars with a visible ``fuzz'' around them.  Thus, the object classes
found by AutoClass are astronomically meaningful---even though the
program itself does not know about stars, galaxies and such!
Moreover, the two morphologically distinct classes of galaxies
populate different regions of the data space, and have systematically
different colors and concentration indices, even though AutoClass was
not given the color information.  Thus, {\it the program has found
astrophysically meaningful distinction between these classes of
objects, which is then confirmed by independent data.}

One critical point in constructing scientifically useful object
catalogs from sky surveys is the classification of astronomical
sources into either stars or galaxies.  Various supervised
classification schemes can be used for this task, including decision
trees~\citep[see, \eg][]{weir95} or neural nets~\citep{odewahn92}.  A more
difficult problem is to provide at least rough morphological types for
the galaxies detected, in a systematic and objective way, without
visual inspection of the images, which is obviously impractical. This
actually provides an interesting opportunity---the application of
new clustering analysis and unsupervised classification techniques may
divide the parent galaxy population into astronomically meaningful
morphological types on the basis of the data themselves, rather than
some preconceived, human-imposed scheme.

\begin{figure}[!hbt]
\plottwo{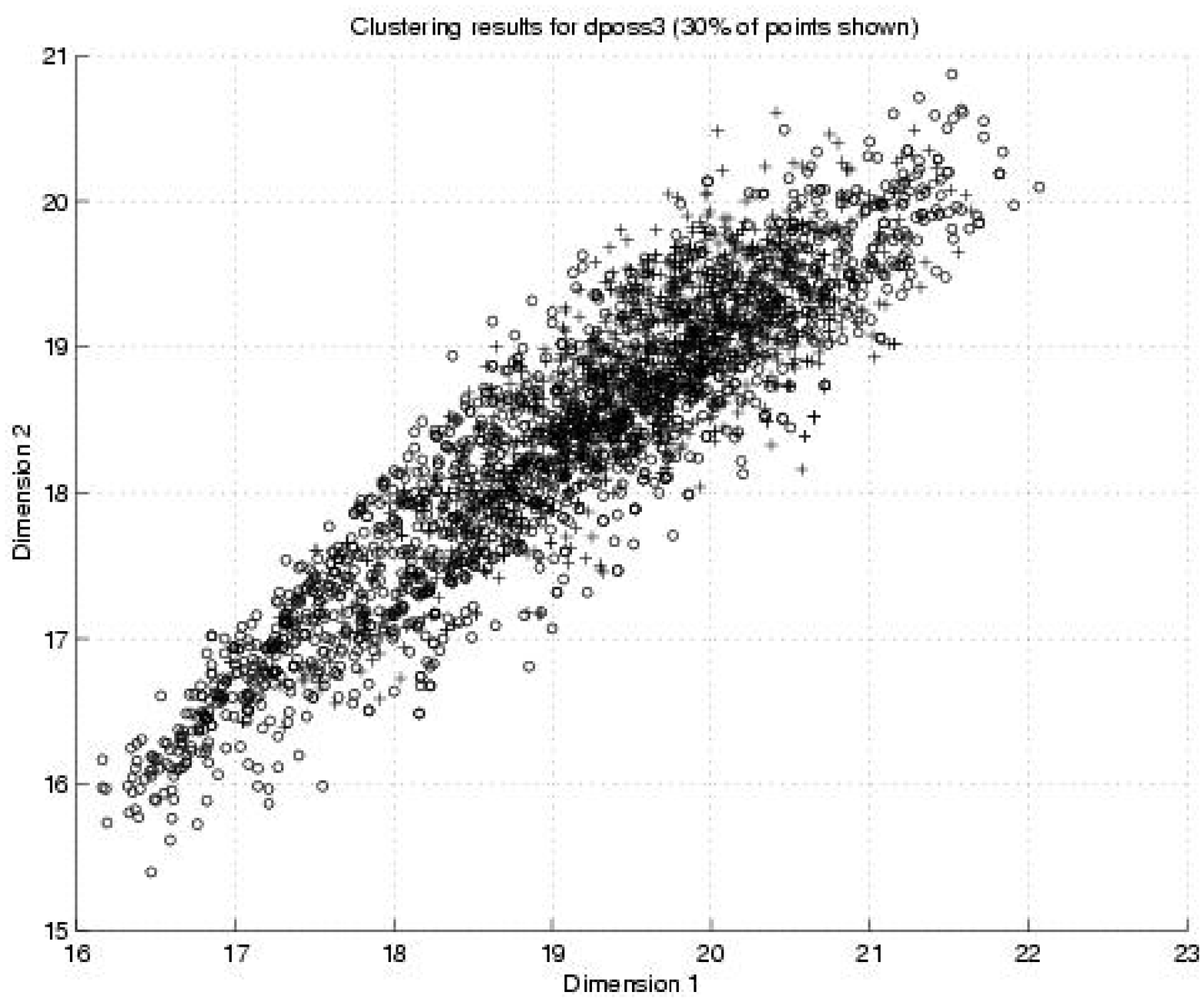}{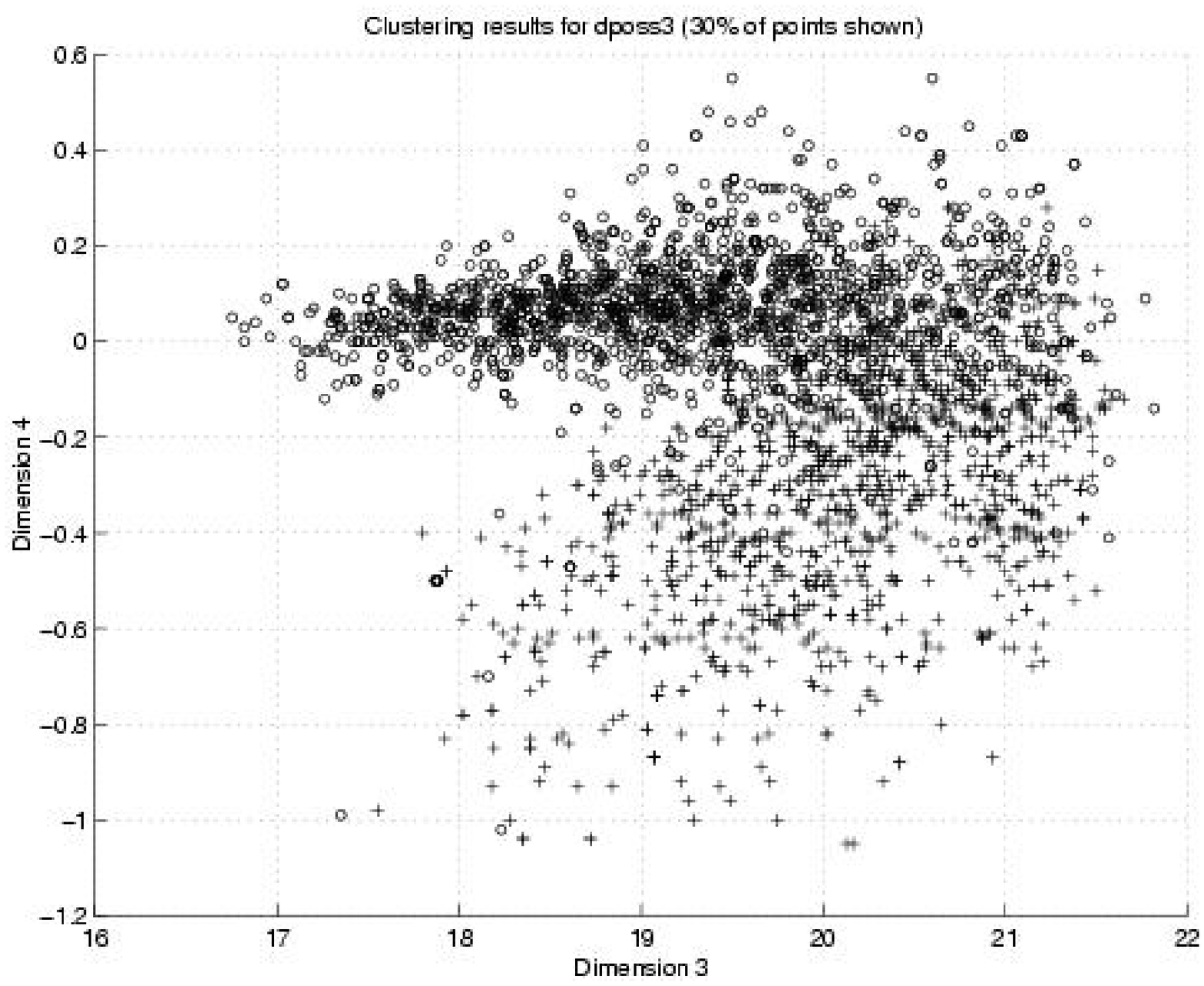}
\caption{An example of unsupervised classification of objects in the
star-galaxy training data set from DPOSS, from the experiment using the
Expectation Maximization multivariate Gaussians mixture modeling.
Two dominant clusters found are shown, encoded as circles (in reality 
stars) and crosses (galaxies).  The plot on the left shows a typical
parameter projection in which the two classes are completely blended.
The plot on the right shows one of the data projections in which the
two classes separate.
\label{emc}}
\end{figure}

Another demonstration of the utility of these techniques can be seen
in Figure~\ref{emc}. In this experiment, the Expectation Maximization
technique was applied on a star-galaxy training data set of
approximately 11,300 objects with 15 parameters each.  This is an
unsupervised classification method which fits a number of multivariate
Gaussians to the data, and decides on the optimal number of clusters
needed to describe the data.  Monte-Carlo cross validation was used to
decide on the optimal number of clusters~\citep[see, \eg][]{smyth00}.  The
program found that there are indeed two dominant classes of objects,
\viz, stars and galaxies, containing about 90\% of all objects, but
that there are also a half-dozen other significant clusters, most of
which correspond to legitimate subclasses such as saturated stars,
\etc Again, this illustrates the potential of unsupervised
classification techniques for objective partitioning of data,
identification of artifacts, and possibly even discovery of new
classes of objects.

\subsubsection{Analyzing Large, Complex Datasets}

Most clustering work in the past has only been applied to small data
sets. The main reasons for this are due to memory storage and
processing speed. With orders of magnitude improvement in both, we can
now begin to contemplate performing clustering on the large scale.
However, clustering algorithms have high computational complexity
(from high polynomial, order 3 or 4, to exponential search). Hence a
rewriting of these algorithms, shifting the focus from performing
expensive searches over small data sets, to robust (computationally
cheap) estimation over very large data sets is in order.

The reason we need to use large data sets is motivated by the fact
that new classes to be discovered in the data are likely to be rare
occurrences (else humans would have surely found them).  For example,
objects like quasars (extremely luminous, very distant sources that
are believed to be powered by supermassive black holes) constitute a
tiny proportion of the huge number of objects detectable in our
survey, yet they are an extremely important class of objects.  Unknown
types (classes) of objects that may potentially be present in data are
likely to be as rare.  Hence, if an automated discovery algorithm is
to have any hope of finding them, it must be able to process a huge
amount of data, millions of objects or more.

Current clustering algorithms simply cannot run on more than a few
thousand cases in less than 10--dimensional space, without requiring
weeks of CPU time.

\begin{enumerate}

\item Many clustering codes (\eg AutoClass) are written to demonstrate 
the method, and are ill-suited for data sets containing millions or
billions of data vectors in tens of dimensions. Improving the
efficiency of these algorithms as the size and complexity of the
datasets is increased is an important issue.

\item With datasets of this size and complexity, multi-resolution 
clustering is a must.  In this regime, expensive parameters to
estimate, such as the number of classes and the initial broad
clustering are quickly estimated using traditional techniques like
K-means clustering or other simple distance-based
method~\citep{duda81}. With such a clustering one would proceed to
refine the model locally and globally.  This involves iterating over
the refinements until some objective (like a Bayesian criterion) is
satisfied.

\item Intelligent sampling methods where one forms ``prototypes''of the
case vectors and thus reduces the number of cases to
process. Prototypes can be determined based on nearest-neighbor type
algorithms or K-means to get a rough estimate, then more sophisticated
estimation techniques can refine this.  A prototype can represent a
large population of examples.  A weighting scheme based on number of
cases represented by each prototype, as well as variance parameters
attached to the feature values assigned to the prototype based on
values of the population it represents, are used to describe them.  A
clustering algorithm can operate in prototype space.  The clusters
found can later refined by locally replacing each prototype by its
constituent population and reanalyzing the cluster.

\item Techniques for dimensionality reduction, including principal
component analysis and others can be used as preprocessing techniques
to automatically derive the dimensions that contain most of the
relevant information.  See, \eg the singular-valued decomposition
scheme to find the eigenvectors dominant in the data set in a related
application involving finding small volcanos in Magellan images of
Venus~\citep{fayyad95}.

\item Scientific verification and evaluation, testing, and follow-up 
on any of the newly discovered classes of objects, physical clusters
discovered by these methods, and other astrophysical analysis of the
results.  This is essential in order to demonstrate the actual
usefulness of these techniques for the NVO.

\end{enumerate}

\subsubsection{Scientific Verification}

Testing of these techniques in a real-life data environment, on a set of
representative science use cases, is essential to validate and improve their
utility and functionality.  Some of the specific scientific verification
tests may include:

\begin{enumerate}

\item A novel and powerful form of the quality control for our data products,
as multidimensional clustering can reveal subtle mismatch patterns
between individual sky survey fields or strips, \eg due to otherwise
imperceptible calibration variations.  This would apply to virtually
any other digital sky survey or other patch-wise collated data sets.
Assured and quantified uniformity of digital sky surveys data is
essential for many prospective applications, \eg studies of the
large-scale structure in the universe, \etc

\item A new, objective approach to star-galaxy separation could 
overcome the restrictions of the current accuracies of star-galaxy
classifications that effectively limits the scientific applications of
any sky survey catalog.  Related to this is an objective, automated,
multi-wavelength approach to morphological classification of galaxies,
\eg quantitative typing along the Hubble sequence, or one of the more
modern, multidimensional classification schemes.

\item An automated search for rare, but known classes of objects, through
their clustering in the parameter space.  Examples include high-redshift
quasars, brown dwarfs, or ultraluminous dusty galaxies.

\item An automated search for rare \emph{and as yet unknown} classes 
of astronomical objects or phenomena, as outliers or sparse clusters 
in the parameter space, not corresponding to the known types of objects. 
They would be found in a systematic way, with fully quantifiable 
selection limits. 

\item Objective discovery of clusters of stars or galaxies in the physical
space, by utilizing full information available in the surveys.  This 
should be superior to most of the simple density-enhancement type
algorithms now commonly used in individual surveys.

\item A general, unbiased, multiwavelength search for AGN, and 
specifically a search for the long-sought population of Type 2 quasars.  
A discovery of such a population would be a major step in our 
understanding of the unification models of AGN, with consequences for 
many astrophysical problems, \eg the origins of the cosmic x-ray background.

\end{enumerate}

This clustering analysis would be performed in the (reduced)
measurement space of the catalogs. But suppose the clustering algorithm
picks out a persistent pattern, \eg a set of objects, that for reasons
not obvious to human from the measurements, are consistently clustered
separately from the data.  The next step is for the astronomer to
examine actual survey images to study this class further to verify
discovery or explain scientifically why the statistical algorithms
find these objects different.

Some enhanced tools for image processing, in particular, probabilistic
methods for segmentation (region growing) that are based both on pixel
value, adjacency information, and the prior expectation of the
scientist will need to be used to aid in analysis and possibly
overcome some loss of information incurred when global image
processing was performed.

There are also potential applications of interest for the searches for
Earth-crossing asteroids, where a substantial portion of the sky would
be covered a few times per night, every night.  The addition of the
time dimension in surveys with repeated observations such as these,
would add a novel and interesting dimension to the problem.  While
variable objects obviously draw attention to themselves (\eg
supernovae, gamma-ray bursts, classical pulsating variables, {\em etc.}),
the truth is that we know very little about the variability of the
deep sky in general, and a systematic search for variability in large
and cross-wavelength digital sky archives is practically guaranteed to
bring some new discoveries.

\subsection{Astronomical Data Visualization}

Effective and powerful data visualization would be an essential part
of any virtual observatory. The human eye and brain are remarkably
powerful in pattern recognition, and selection of interesting
features. The technical challenge here is posed by the sheer size of
the datasets (both in the image and catalog domains), and the need to
move through them quickly and to interact with them ``on the fly''.

The more traditional aspect of this is the display of various
large-format survey images.  One of the new challenges is in streaming
of the data volume itself, now already in the multi-TB range for any
given survey.  A user may need to shift quickly through different
spatial scales (\ie zoom in or out) on the display, from the entire
sky down to the resolution limit of the data, spanning up to a factor
of approximately $10^{11}$ (!) in solid angle coverage.  Combining the image
data from different surveys with widely different spatial resolutions
poses additional challenges.  So does the co-registration of images
from different surveys where small, but always present systematic
distortions in astrometric solutions must be corrected before the
images are overlaid.

Another set of challenges is presented by displaying the information
in the parameter spaces defined in the catalog domain, where each
object may be represented by a data vector in tens or even hundreds of
dimensions, but only a few can be displayed at any given time (\eg 3
spatial dimensions, color, shape, and intensity for displayed
objects).  Each of the object attributes, or any user defined
mathematical combination of object attributes (\eg colors) should be
encodeable on demand as any of the displayed dimensions.  This
approach will also need to be extended to enable the display of data
from more than one survey at a time, and to combine object attributes
from matched catalogs.

However, probably the most interesting and novel aspect is the
combination and interaction between the image and catalog domains.
This is only becoming possible now, due to the ability to store
multi-TB data sets on line, and it opens a completely new territory.
In the simplest approach this would involve marking or overplotting of
sources detected in one survey, or selected in some manner, \eg in
clustering analysis, on displayed images.

\begin{figure}[!hbt]
\plotfiddle{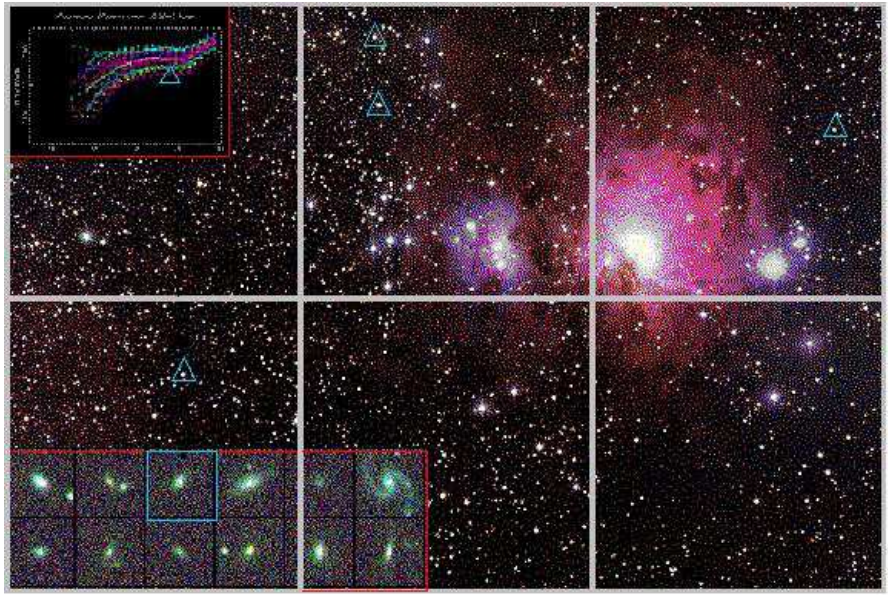}{3.0in}{0}{60}{60}{-172}{0}
\caption{A prototype of the visualization services which would empower scientists 
to not only tackle current scientific challenges, but also to aid in
the exploration of the, as yet unknown, challenges of the future. Note
the intelligent combination of image and catalog visualizations to aid
the scientist in exploring parameter space.Image Credit: Joe Jacob and the ALPHA group,
(JPL).\label{proto-vis}}
\end{figure}

In the next level of functionality, the user would be able to mark the
individual sources or delineate areas on the display, and retrieve the
catalog information for the contained sources from the catalog domain
(see, \eg Figure~\ref{proto-vis} for a demonstration).  Likewise, it
may be necessary to remeasure object parameters in the pixel domain
and update or create new catalog entries.  An example may be measuring
of low-level signals or upper limits at locations where no
statistically significant source was cataloged originally, but where
a source detection is made in some other survey, \eg faint optical
counterparts of IR, radio, or x-ray sources.  An even more
sophisticated approach may involve development of new object
classifiers through interaction of catalog and image domains, \eg
using genetic algorithms.

\begin{figure}[!hbt]
\plotfiddle{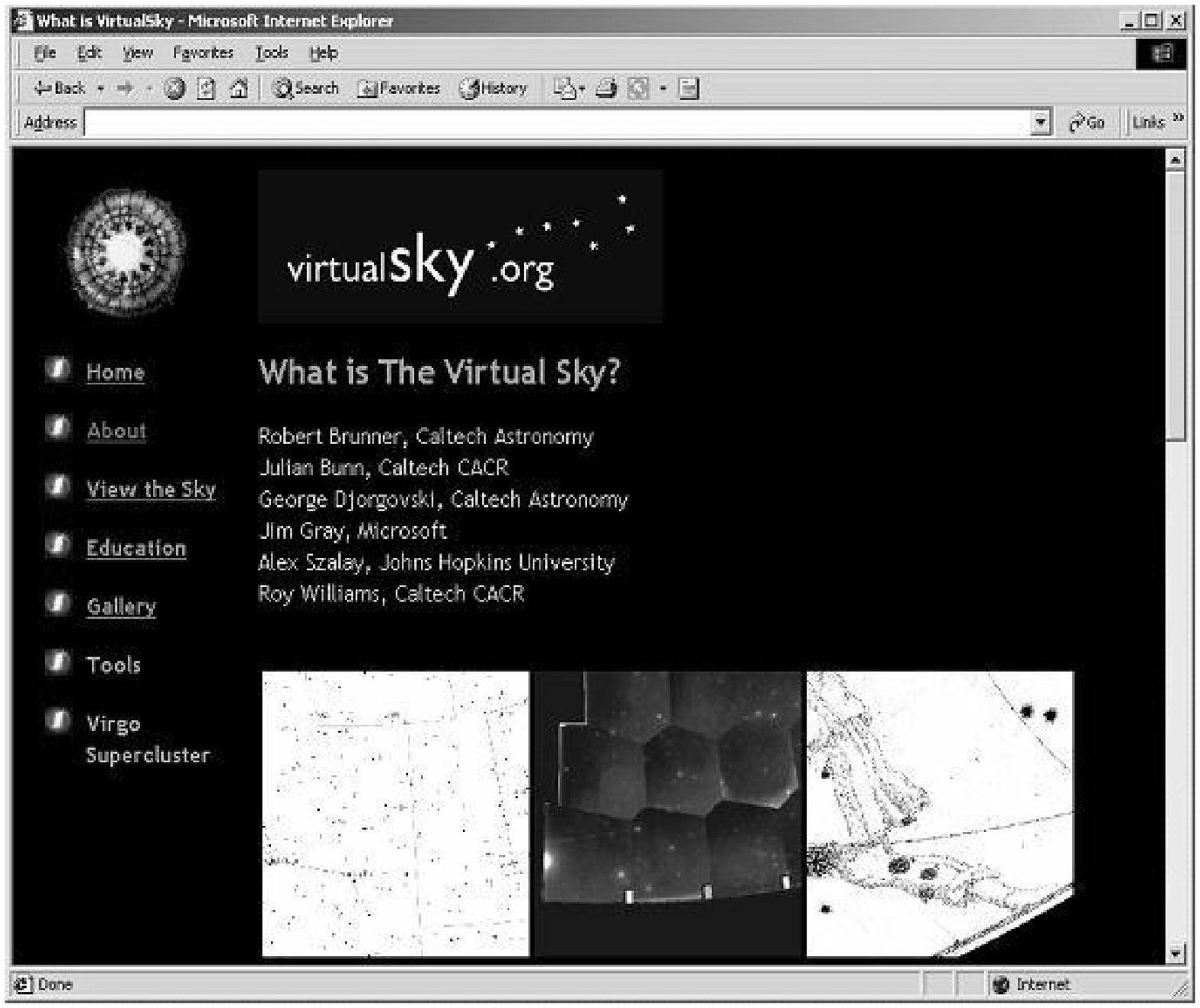}{3.0in}{0}{50}{50}{-160}{-9}
\caption{The Virtual Sky web-site which provides public access to several different data products, 
including images from the DPOSS survey.\label{vo}}
\end{figure}

Visualization of these large digital sky surveys is also a powerful
education and public outreach tool. An example of this is the virtual
sky project\footnote{http://www.virtualsky.org} (see Figure~\ref{vo}
for the project homepage).

\section{Summary}

We are at the start of a new era of information-rich astronomy.
Numerous ongoing sky surveys over a range of wavelengths are already
generating data sets measured in the tens of Terabytes.  These surveys
are creating catalogs of objects (stars, galaxies, quasars, {\em etc.})
numbering in billions, with tens or hundreds of measured numbers for
each object.  Yet, this is just a foretaste of the much larger data
sets to come, with multi-Petabyte data sets already on the horizon.
Large digital sky surveys and data archives are thus becoming the principal
sources of data in astronomy.  The very style of observational
astronomy is changing: systematic sky surveys are now used both to
answer some well-defined questions which require large samples of
objects, and to discover and select interesting targets for follow-up
studies with space-based or large ground-based telescopes.

This vast amount of new information about the universe will enable and
stimulate a new way of doing astronomy.  We will be able to tackle
some major problems with an unprecedented accuracy, \eg mapping of
the large-scale structure of the universe, the structure of our
Galaxy, \etc The unprecedented size of the data sets will enable
searches for extremely rare types of astronomical objects (\eg
high-redshift quasars, brown dwarfs,  {\em etc.}) and may well lead to
surprising new discoveries of previously unknown types of objects or
new astrophysical phenomena.  Combining surveys done at different
wavelengths, from radio and infrared, through visible light,
ultraviolet, and x-rays, both from the ground-based telescopes and
from space observatories, would provide a new, panchromatic picture of
our universe, and lead to a better understanding of the objects in
it. These are the types of scientific investigations which were not
feasible with the more limited data sets of the past.

Many individual digital sky survey archives, servers, and digital
libraries already exist, and represent essential tools of modern
astronomy.  We have reviewed some of them, and there are many others
existing and still under development.  However, in order to join or
federate these valuable resources, and to enable a smooth inclusion of
even greater data sets to come, a more powerful infrastructure and a
set of tools are needed.

The concept of a virtual observatory thus emerged, including the
incipient National Virtual Observatory (NVO), and its future global
counterparts.  A virtual observatory would be a set of federated,
geographically distributed, major digital sky archives, with the
software tools and infrastructure to combine them in an efficient and
user-friendly manner, and to explore the resulting data sets whose
sheer size and complexity are beyond the reach of traditional
approaches.  It would help solve the technical problems common to most
large digital sky surveys, and optimize the use of our resources.

This systematic, panchromatic approach would enable new science, in
addition to what can be done with individual surveys. It would enable
meaningful, effective experiments within these vast data parameter
spaces. It would also facilitate the inclusion of new massive data
sets, and optimize the design of future surveys and space missions.
Most importantly, the NVO would provide access to powerful new
resources to scientists and students everywhere, who could do
first-rate observational astronomy regardless of their access to large
ground-based telescopes.  Finally, the NVO would be a powerful
educational and public outreach tool.

Technological challenges inherent in the design and implementation of
the NVO are similar to those which are now being encountered in other
sciences, and offer great opportunities for multi-disciplinary
collaborations.  This is a part of the rapidly changing,
information-driven scientific landscape of the new century.

\begin{acknowledgments}
We would like to acknowledge our many collaborators in the Virtual
Observatory initiative. Particular thanks go to the members of the
Digital Sky project and the NVO interim steering committee, where many
of the ideas presented herein were initially discussed. We would also
like to thank the editors for the immense patience with the
authors. This work was made possible in part through the NPACI
sponsored Digital Sky Project (NSF Cooperative Agreement ACI-96-19020)
and generous grants from the SUN Microsystems Corporation and
Microsoft Research. RJB would like to personally acknowledge financial
support from the Fullam Award and NASA grant number NAG5-10885. SGD
acknowledges support from NASA grants NAG5-9603 and NAG5-9482, the
Norris foundation, and other private donors.
\end{acknowledgments}

\bibliographystyle{apalike}
\bibliography{brunnerr}

\begin{thebibliography}{}

\bibitem[{Aubele} et~al., 1995]{fayyad95}
{Aubele}, J.~C., {Crumpler}, L.~S., {Fayyad}, U.~M., {Smyth}, P., {Burl},
  M.~C., and {Perona}, P. (1995).
\newblock Locating small volcanoes on venus using a scientist-trainable
  analysis system.
\newblock In {\em Lunar and Planetary Science Conference}, volume~26, page~61.

\bibitem[Brunner et~al., 2001]{NVO2K}
Brunner, R.~J., Djorgovski, S.~G., and Szalay, A.~S., editors (2001).
\newblock {\em {Virtual Observatories of the Future}}. Astronomical Society of
  the Pacific, San Francisco.

\bibitem[Cheeseman et~al., 1988]{cheeseman88}
Cheeseman, P., Kelly, J., Self, M., Stutz, J., Taylor, W., and Freeman, D.
  (1988).
\newblock {AutoClass: A Bayesian Classification System}.
\newblock In {\em {Proceedings of the Fifth International Conference on Machine
  Learning}}, pages {54--64}. Morgan Kaufmann Publishers, San Francisco.

\bibitem[{de Carvalho} et~al., 1995]{decarvalho95}
{de Carvalho}, R.~R., {Djorgovski}, S.~G., {Weir}, N., {Fayyad}, U.,
  {Cherkauer}, K., {Roden}, J., and {Gray}, A. (1995).
\newblock Clustering analysis algorithms and their applications to digital
  poss-ii catalogs.
\newblock In {\em Astronomical Data Analysis Software and Systems IV},
  volume~77 of {\em ASP Conference Series}, pages {272--275}. Astronomical
  Society of the Pacific, San Francisco.

\bibitem[Duda and Hart, 1981]{duda81}
Duda, R. and Hart, P. (1981).
\newblock {\em Pattern Classification and Scene Analysis}.
\newblock John Wiley, \& Sons.

\bibitem[Foster and Kesselman, 2001]{grid}
Foster, I. and Kesselman, C., editors (2001).
\newblock {\em The Grid: Blueprint for a New Computing Infrastructure}.
\newblock Morgan Kaufmann, San Francisco, CA.

\bibitem[Lonsdale et~al., 1998]{lonsdale98}
Lonsdale, C.~J., Conrow, T., Evans, T., Fullmer, L., Moshir, M., Chester, T.,
  Yentis, D., Wolstencroft, R., MacGillivray, H., and Egret, D. (1998).
\newblock {The OPTID Database: Deep Optical Identifications to the IRAS Faint
  Source Survey}.
\newblock In McLean, B.~J., Golombek, D.~A., Hayes, J. J.~E., and Payne, H.~E.,
  editors, {\em New Horizons from Multi-Wavelength Sky Surveys}, volume 179 of
  {\em IAU Symposium Series}, pages 450--451. Kluwer Academic Publishers,
  Netherlands.

\bibitem[{McMahon} and {Irwin}, 1992]{mcmahon92}
{McMahon}, R.~G. and {Irwin}, M.~J. (1992).
\newblock Apm surveys for high redshift quasars.
\newblock In {\em ASSL Vol. 174: Digitised Optical Sky Surveys}, page 417.

\bibitem[Monet et~al., 1996]{monet96}
Monet, D., Bird, A., Canzian, B., Harris, H., Reid, N., Rhodes, A., Sell, S.,
  Ables, H., Dahn, C., Guetter, H., Henden, A., Leggett, S., Levison, H.,
  Luginbuhl, C., Martini, J., Monet, A., Pier, J., Riepe, B., Stone, R., Vrba,
  F., and Walker, R. (1996).
\newblock {USNO-SA2.0}.
\newblock Technical report, U.S. Naval Observatory, Washington DC.

\bibitem[{NVO Informal Steering Committee}, 2001]{nvo-white}
{NVO Informal Steering Committee} (2001).
\newblock Towards a national virtual observatory: Science goals, technical
  challenges, and implementation plan.
\newblock In Brunner, R., Djorgovski, S., and Szalay, A., editors, {\em Virtual
  Observatories of the Future}, pages 353--372. Astronomical Society of the
  Pacific, San Francisco.

\bibitem[{Odewahn} et~al., 1992]{odewahn92}
{Odewahn}, S.~C., {Stockwell}, E.~B., {Pennington}, R.~L., {Humphreys}, R.~M.,
  and {Zumach}, W.~A. (1992).
\newblock Automated star/galaxy discrimination with neural networks.
\newblock {\em Astronomical Journal}, 103:318--331.

\bibitem[{Pennington} et~al., 1993]{pennington93}
{Pennington}, R.~L., {Humphreys}, R.~M., {Odewahn}, S.~C., {Zumach}, W., and
  {Thurmes}, P.~M. (1993).
\newblock The automated plate scanner catalog of the palomar sky survey. i -
  scanning parameters and procedures.
\newblock {\em Publications of the Astronomical Society of the Pacific},
  105:521--526.

\bibitem[{Reid} et~al., 1991]{reid91}
{Reid}, I.~N., {Brewer}, C., {Brucato}, R.~J., {McKinley}, W.~R., {Maury}, A.,
  {Mendenhall}, D., {Mould}, J.~R., {Mueller}, J., {Neugebauer}, G., {Phinney},
  J., {Sargent}, W. L.~W., {Schombert}, J., and {Thicksten}, R. (1991).
\newblock The second palomar sky survey.
\newblock {\em Publications of the Astronomical Society of the Pacific},
  103:661--674.

\bibitem[{Rutledge} et~al., 2000]{rutledge2k}
{Rutledge}, R.~E., {Brunner}, R.~J., {Prince}, T.~A., and {Lonsdale}, C.
  (2000).
\newblock {XID: Cross-Association of ROSAT/Bright Source Catalog X-Ray Sources
  with USNO A-2 Optical Point Sources}.
\newblock {\em Astrophysical Journal, Supplement}, 131:{335--353}.

\bibitem[Smyth, 2000]{smyth00}
Smyth, P. (2000).
\newblock Model selection for probabilistic clustering using cross-validated
  likelihood.
\newblock {\em Statistics and Computing}, 10:63--72.

\bibitem[Szalay and Brunner, 1999]{fgcs}
Szalay, A. and Brunner, R.~J. (1999).
\newblock {Astronomical Archives of the Future: A Virtual Observatory}.
\newblock {\em {Future Generations of Computational Systems}}, 16:63.

\bibitem[{Urban} et~al., 1997]{urban97}
{Urban}, S., {Corbin}, T., and {Wycoff}, G. (1997).
\newblock The act reference catalog.
\newblock In {\em American Astronomical Society Meeting}, volume 191, page
  5707.

\bibitem[{Voges} et~al., 1999]{voges99}
{Voges}, W., {Aschenbach}, B., {Boller}, T., {Br{\"a}uninger}, H., {Briel}, U.,
  {Burkert}, W., {Dennerl}, K., {Englhauser}, J., {Gruber}, R., {Haberl}, F.,
  {Hartner}, G., {Hasinger}, G., {K{\"u}rster}, M., {Pfeffermann}, E.,
  {Pietsch}, W., {Predehl}, P., {Rosso}, C., {Schmitt}, J. H. M.~M.,
  {Tr{\"u}mper}, J., and {Zimmermann}, H.~U. (1999).
\newblock The rosat all-sky survey bright source catalogue.
\newblock {\em Astronomy and Astrophysics}, 349:389--405.

\bibitem[Weir et~al., 1995]{weir95}
Weir, N., Fayyad, U., Djorgovski, S.~G., and Roden, J. (1995).
\newblock {The SKICAT System for Processing and Analyzing Digital Imaging Sky
  Surveys}.
\newblock {\em Publications of the Astronomical Society of the Pacific},
  107:1243.

\bibitem[{White} et~al., 1994]{white94}
{White}, N.~E., {Giommi}, P., and {Angelini}, L. (1994).
\newblock Wgacat.
\newblock {\em IAU Circular}, 6100:1.

\bibitem[{Yentis} et~al., 1992]{yentis92}
{Yentis}, D.~J., {Cruddace}, R.~G., {Gursky}, H., {Stuart}, B.~V., {Wallin},
  J.~F., {MacGillivray}, H.~T., and {Collins}, C.~A. (1992).
\newblock The cosmos/ukst catalog of the southern sky.
\newblock In {\em ASSL Vol. 174: Digitised Optical Sky Surveys}, page~67.

\end{thebibliography}

\end{document}